\documentclass[prc,aps]{revtex4-1}
\usepackage{graphicx}
\usepackage{natbib}
\usepackage{xcolor}

\usepackage{subfig}
\usepackage{float}
\usepackage{amsmath}
\usepackage{appendix}
\begin{document}

\title{A modified Brink-Axel hypothesis for astrophysical Gamow-Teller transitions}

\author{Ra\'ul A. Herrera}
\affiliation{Department of Physics and \\
University of California San Diego,
9500 Gilman Drive, La Jolla, CA 92093} 
\author{Calvin W. Johnson}
\affiliation{Department of Physics, San Diego State University,
5500 Campanile Drive, San Diego, CA 92182}
\author{George M. Fuller}
\affiliation{Department of Physics and \\
University of California San Diego,
    9500 Gilman Drive, La Jolla, CA 92093 }

\begin{abstract}
Weak interaction charged current transition strengths from highly excited nuclear states are 
fundamental ingredients for accurate modeling of compact object composition and dynamics, 
but are difficult to obtain either from experiment or theory.  For lack of alternatives, 
calculations have often fallen back upon a generalized Brink-Axel hypothesis, that is, assuming  the strength function 
(transition probability) 
is independent of the initial nuclear state but depends only upon the transition energy and the weak interaction properties of the parent nucleus ground state.
Here we present numerical evidence for a modified `local' Brink-Axel hypothesis for Gamow-Teller transitions for  $pf$-shell nuclei   relevant to astrophysical applications. Specifically, while the 
original Brink-Axel hypothesis does not hold globally, strength functions from initial states nearby in 
energy are  similar within statistical fluctuations. This agrees with previous work 
on strength function moments. Using this modified hypothesis, we can tackle strength functions at 
previously intractable initial energies, using semi-converged initial states  at arbitrary excitation energy.
Our work provides a well-founded method  for computing  accurate thermal weak transition rates  for medium-mass nuclei  at temperatures occurring in stellar cores near collapse. 
We finish by comparing to previous calculations of astrophysical rates.
\end{abstract}

\maketitle

\section{Introduction}

Despite its unassuming moniker, the weak nuclear force is a prime driver of stellar evolution. It facilitates nuclear energy generation and the long ($\approx 10\,{\rm Gyr}$) hydrogen-burning lives for stars like the sun, but it is also key to understanding some of the most violent astrophysical events in the universe~\cite{FowlerHoyle64,betheGAL79,Fuller1982a,CoopWambach1984,Bethe1990,Burrows2013,janka2007theory,MassiveStars}. In fact, a cabal of the weak interaction and gravitation join forces to bring down stars with masses considerably greater than that of the sun. The evolution of stars with initial masses greater than $\approx 8\,{\rm M}_\odot$ is dominated by the weak interaction, specifically by entropy and lepton number loss, in part through nuclear electron capture and positron decay processes. The escaping neutrinos produced by these and other neutrino emission processes \lq\lq refrigerate\rq\rq\ the cores of these massive stars, lowering the entropy to $s \approx 1$, in units of Boltzmann's constant $k$ per baryon, and thereby ensuring that nucleons reside primarily in nuclei and that the pressure support for the star stems from degenerate electrons with relativistic kinematics. Self-gravitating configurations are trembling on the verge of instability whenever they are supported by the pressure of particles moving near the speed of light. Collapse of the cores of these stars is inevitable. 

The detailed history of nuclear weak transition processes like electron capture and positron decay in the pre-collapse evolution of these massive 
stars determines the entropy, temperature, and lepton fraction that governs the initial mass of the collapsing core\cite{Bethe1990,heger2001presupernova}. In fact, not only are nuclear weak interaction processes instrumental in determining the thermodynamic conditions and composition at the onset of collapse, they are also important during the collapse itself, largely determining the mean nuclear mass, and electron lepton fraction and entropy generation. Despite the low entropy of the pre-collapse configurations, the temperature at the onset of collapse can be high, with $kT \approx 1\, {\rm MeV}$, implying that the baryonic component will be in nuclear statistical equilibrium (NSE) and comprised of iron peak nuclei. As the core collapses and density rises, and the temperature rises modestly, NSE shifts to nuclei of even higher mass number. These heavier nuclei also have higher neutron excess, reflecting the cumulative effect of electron capture and electron lepton number loss as neutrinos (at first) escape.

The predominant weak interaction process near and during collapse is electron capture on protons, both free protons and protons inside these heavy nuclei. The weak nuclear matrix elements for electron capture on free protons far exceed the effective \lq\lq per-proton\rq\rq\ matrix elements for protons inside nuclei. Consequently, the overall rate of electron capture, the neutronization rate, depends sensitively on the mass fraction in free protons. In turn, that quantity depends sensitively on temperature and entropy. The primary entropy source during collapse is electron capture on nuclei. This process leaves a hole in the degenerate electron energy distribution, plus the weak interaction selection rules and strength function characteristics favor leaving the daughter nucleus in an extra-thermal excited state. Both of these outcomes represent an increase in entropy-per-baryon. Higher entropy favors a higher free proton fraction and so faster neutronization. This entropy production and free proton fraction feedback loop plays out during collapse against a backdrop of electron capture- and density rise-induced increase in mean nuclear mass. That increase eventually leads to the trapping of neutrinos through neutral current neutrino coherent scattering on nuclei and to rapid subsequent thermalization of the neutrinos through neutrino-electron scattering and through inelastic neutrino-nucleus scattering \cite{FullerMeyer91,Langanke2008,Langanke21,Wendell2013,Fischer2013}. The electron fraction (number of electrons per baryon) $Y_e$ is determined by the integrated history of electron capture and lepton number loss through escaping neutrinos. The mass of the homologous core scales like $Y_e^2$. This core serves as the \lq\lq piston\rq\rq\ that drives the initial supernova shock with an energy that scales like $Y_e^{10/3}$. That shock losses energy and eventually stalls at a radius $\approx 200\,{\rm km}$, but can be re-energized through neutrinos radiated by the hot proto-neutron star and captured in the matter behind the shock, all aided by hydrodynamic transport \cite{janka2007theory,MassiveStars}. 
{In addition, it has been recently argued that the neutrino radiance from nearby stars may give us advance warning of a core-collapse supernova~\cite{patton2017neutrinos}.}
All of these considerations highlight the central role of nuclear weak interactions in core collapse and in compact object composition, dynamics, and nucleosynthesis \cite{Hix2003}.


As a consequence,
researchers need reliable weak nuclear transition probabilities, also called transition strengths. 
From experiment one 
either has data from beta decays and electron captures, primarily although not exclusively from the ground state, 
or allowed transitions strengths, again from the ground state, through $(p,n)$ and $(n,p)$ type reactions~\cite{doering1975observation,PhysRevLett.44.1755}. 
But extremes of temperature and pressure mean the nuclei in the cores of massive starts can be 
excited by many MeV, so that astrophysical calculations require many transitions where both the initial and final 
states are highly excited.  Not only are such transitions difficult or impossible to access experimentally, 
they are also challenging to model theoretically.

What does one do then for weak transitions between highly excited states?  The most frequent 
resolution is to invoke a generalization of the Brink-Axel hypothesis \cite{FFNII} for Gamow-Teller (axial vector) transitions, inspired by an analogy to the distribution of Fermi vector weak transition strength to isobaric analog states and by an analogy to electromagnetic collective modes. 

Originally Brink~\cite{BrinkThesis}, and independently Axel~\cite{PhysRev.126.671},  
hypothesized that giant electric dipole resonances built upon excited states should resemble the 
giant electric dipole resonance  built upon the ground state.  This hypothesis arose in part 
from the picture of an electric dipole resonance arising from the nuclear proton component and the neutron component collectively oscillating against one another, with the argument that such a simple picture 
should be insensitive to the detailed structure of the initial state.  Later the Brink-Axel 
hypothesis was generalized to other kinds of collective nuclear transitions, in part because of the lack of practical  
alternatives. Given its lack of rigor, the  generalized Brink-Axel hypothesis  has remained 
controversial (see the discussion in ~\cite{SumRulesBrink}). 

Building upon prior work, in this paper  we provide strong numerical evidence for a modified Brink-Axel hypothesis for
Gamow-Teller strength functions: we find strength functions from initial states nearby in energy 
are statistically similar. 
This ``energy-localized'' Brink-Axel hypothesis allows us to extract useful strength functions from states with  high initial energies. 
Because strength functions from initial states close in energy are similar, we don't need fully converged initial states.
Instead we can use semi-converged  states, which are superpositions of states nearby in energy, as proxies for fully converged initial states, and from them 
generate statistically typical strength functions.  

In the next section we discuss the astrophysical context for  weak transitions, followed by 
a review of the 
Brink-Axel hypothesis and a brief overview 
of prior calculations of astrophysical weak transitions in medium-mass nuclides, especially around the crucial iron region.
After outlining how one computes strength functions from realistic shell model calculations, 
we provide evidence for our energy-localized Brink-Axel hypothesis (ELBAH) for cases where we can extract strength functions 
from fully converged states up to high energy.  In particular, we use running sums of Gamow-Teller strengths and take 
binned averages; we find that the fluctuations are not very sensitive to the size of the initial energy bin, which we 
interpret as meaning that we can usefully treat the strength functions as statistically similar.  Building upon 
this, we use a novel thick-restart Lanczos algorithm, described in the Appendix, which produces semi-converged states at high energy and 
extract Gamow-Teller strength functions for several nuclides. The strength function themselves are also computed using 
another variant of the Lanczos algorithm.  Finally we use the ELBAH to generate astrophysical 
rates for $^{57}$Co$\rightarrow ^{57}$Fe  and compare to prior calculations.

\section{Background: The astrophysical context of weak transitions}

\label{background}

As is well understood, a massive star's prodigious required pressure support against gravitation necessitates  high temperatures and densities. The high temperatures help overcome Coulomb barriers and enable 
 exothermic fusion~\cite{clayton1983principles}.  
 In this work we are concerned with the events leading up to the gravitational collapse of the core of a massive star and how weak nuclear reaction rates affect and accelerate this process~\cite{MassiveStars}. Medium-mass nuclides around iron are especially relevant.


While the core of our Sun has a density of about 150 g/cm$^3$ and a temperature of $15 \times 10^6$ K, a massive star in the final stages 
of its life will have core densities of $10^{8}$-$10^{10} \mathrm{g/cm^3}$ and temperatures up to a few times $10^{10}$K ~\cite{MassiveStars}. This means 
the pre-collapse astrophysical environment has energies in the range from 1 keV to several MeV, and during collapse can be even hotter.
Because in pre-supernova cores the degenerate electron pressure eventually dominates over thermal pressure and determines the 
Chandrasekar mass~\cite{heger2001presupernova}, the density of electrons is crucial to know. This electron density is usually cast in terms of the lepton fraction $Y_e$, the 
ratio of electrons to baryons.


For any weak process the astrophysical rate as a function of temperature $T$ and mass density $\rho$ is
\begin{equation}
\lambda(\rho,T)=\sum_i P_i\sum_f \lambda_{if}=\displaystyle\sum_i\frac{(2J_i+1)e^{-E_i/kT}}{G(T)}\displaystyle\sum_f B_{if}({\cal O})\Phi_{if}(Q_{if},\rho,T).
\label{eqn:thermalrate}
\end{equation}
Here $i$ labels initial levels and $f$ final levels, both of which are eigenstates of the nuclear Hamiltonian with 
energies $E_i, E_f$, respectively.
The partial transition rate from $i$ to $f$,
\begin{equation}
\lambda_{if}= B_{if} \Phi_{if} 
\end{equation}
depends upon the transition probability (or transition strength) $B_{if}$ and the phase-space factor $\Phi_{if}$; we discuss the 
latter  in more detail in section \ref{application}.

The effect of temperature plays out in the initial state thermal occupation probability and in the nuclear
partition function
\begin{equation}
G(T) = \sum_i (2J_i +1) e^{-E-i/kT} ,   
\end{equation}
where $J_i$ is the angular momentum of the initial level $i$ and $k$ is Boltzmann's constant.  Then $P_i=(2J_i+1)e^{-E_i/kT}/G(T)$ is the occupation probability for initial state $i$. 

But at high temperature  nuclei can be highly excited.  How high does one have to go, and how many levels need to be included?
In the Fermi gas approximation~\cite{betheGAL79}, at a temperature $T$  the average excitation energy is 
\begin{equation}
    E_x=a(kT)^2,
    \label{eqn:excite}
\end{equation}
where $a\approx\frac{A}{8 \, \mathrm{MeV}}$ is the level density parameter parameter in terms of the nuclear mass number $A$. For example, if  $A=57$ and $T = 12 \times 10^9$ K or $kT=1$ MeV, then a typical excitation energy $E_x \approx $ 7 MeV.

\begin{figure}
    \centering
    \includegraphics[scale=0.4,clip]{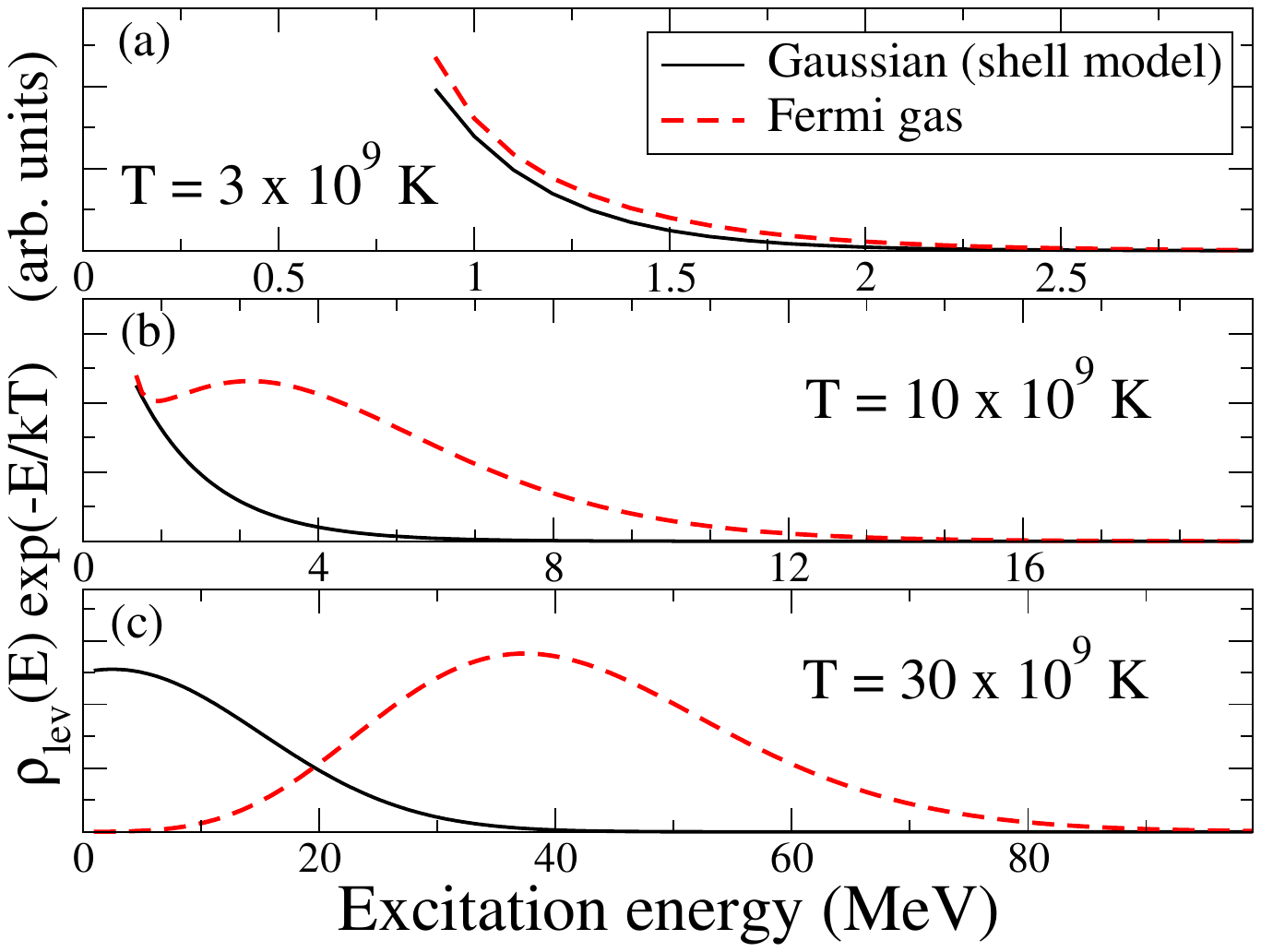}
    \caption{Product of the Boltzmann factor $\exp(-E/kT)$ times the level density $\rho_\mathrm{lev}(E)$ for two models of level densities, 
a Gaussian with parameters taken from the shell model (solid line), and the backshifted Fermi gas model (dashed line), at temperatures 
(a) $= 3 \times 10^9$ K; (b) $= 10 \times 10^9$ K; and (c) (a) $= 30 \times 10^9$ K. Note the different energy scales at different 
temperatures.} 
    \label{fig:levelboltzmann}
\end{figure}




To illustrate  in more detail, consider  the product of the increasing level density $\rho_\mathrm{lev}(E_i)$ and the decreasing Boltzmann factor $\exp(-E_i/kT)$. 
The modeling of the level density is itself a thorny issue, one which we will not resolve here.  We take two models: as a good approximation to the shell model, 
a Gaussian distribution
\begin{equation}
\rho_\mathrm{lev}(E)=\frac{N}{\sigma\sqrt{2\pi}}e^{-\frac{1}{2}(\frac{E-\bar{E}}{\Delta E})^2}, 
\end{equation}
where $N$ is the total number of levels, $\bar{E}$ is the energy centroid, and $\Delta E$ is the width~\cite{mon1975statistical,chang1972validity}; 
and the  back-shifted Fermi gas model,
\begin{equation}
    \rho_\mathrm{lev}(E)=\frac{\exp{\left(2\sqrt{a(E-\delta)}\right)}}{12\sqrt{2}\sigma a^{1/4}(E - \delta)^{5/4}}
\end{equation}
where $a$ is the level density parameter, $\delta$ is the backshift parameter, and $\sigma$ is a spin-cutoff factor~\cite{ericson1960statistical}. 
In section~\ref{casestudy} we make a case study for the thermal rates of $^{57}$Co $\rightarrow ^{57}$Fe, and so here 
we take  specific numbers for $^{57}$Co:
$N=980$ million is the total number of levels, $\Delta E=12.55$ MeV, and $\bar{E}=63.4$ Mev above the ground state for the 
Gaussian shell model level density. For the Fermi gas model we use the parameters 
 $a=6.5$ MeV  and $\delta=0.7$ MeV from  Mishra et al. \cite{MishraGrimesLevel}; as the spin-cutoff parameter $\sigma$ is not given, we use $\sigma=3.96$ from $pf$-shell calculations \cite{Spinella}, which found $\sigma$  to be approximately constant over a wide range of energies.
 
 We plot $\rho_\mathrm{lev}(E)  \times \exp(- E/kT)$ in Fig.~\ref{fig:levelboltzmann} at three different temperatures and for the two models for level 
 densities.  There are two lessons to be learned here. The first is that while up to a few billion degrees Kelvin, see Fig.~~\ref{fig:levelboltzmann}(a),  one will only 
 excite the lowest levels; but once one reaches ten billion degrees Kelvin, Fig.~~\ref{fig:levelboltzmann}(b), one excites levels at a few MeV, and by thirty  billion degrees Kelvin, Fig.~~\ref{fig:levelboltzmann}(c), one can excited to tens of MeV. 
 The other lesson of course is that this depends upon the model of level densities. The shell model space is finite and so 
 the  product $\rho_\mathrm{lev}(E) \exp(- E/kT)$ for the shell model falls off faster than for 
 the backshifted Fermi gas model, which includes levels outside the shell model space; these latter are called `intruders.' 
 Whether strength functions from such intruder states follow the same patterns as the shell model strength functions found below is a question we do not attempt to answer here.
 
 Nonetheless, it is clear that, no matter how one models the level densities, at high astrophysical temperatures one can access 
 highly excited states. How to accurately model the strength functions of these highly excited states is the main topic of this paper.

\subsection{Strength functions and the Brink-Axel hypothesis}

\label{BAHbasics} 

The transition strength or $B$-value is the transition matrix element squared, averaged over initial states 
and summed over final states, and can be written in terms of a reduced matrix element~\cite{edmonds1996angular}, 
\begin{equation}
B_{if}({\cal O})=\frac{1}{2J_i+1}|\langle J_f\, T_f||\hat{\cal O} ||J_i\, T_i\rangle|^2
\label{eqn:BGT}
\end{equation}
where $\hat{\cal O}$ is the transition operator; here we focus on the Gamow-Teller operator $g_A \vec{\sigma} \tau$. 

 To characterize transition strengths, 
one introduces the strength function,
\begin{equation}
    B(E_i,E_\mathrm{tr})=\sum_f \delta (E_\mathrm{tr}-E_f+E_i) B_{if}.
    \label{eqn:strfunc}
\end{equation}
Often this is written as 
$S(E_i,E_\mathrm{tr})$, while $B_{if}$ is used for single transitions, but we will use $B$ for consistency.

A central problem is this: while there exist efficient methods for computing 
the strength function for the ground state with the Lanczos algorithm (see also the Appendix), and while one can in some cases 
measure the strength function off the ground state~\cite{doering1975observation,PhysRevLett.44.1755}, strength functions for highly excited states are generally not achievable 
experimentally and are challenging theoretically. Hence one often turns to  the Brink-Axel hypothesis (BAH), which, put simply, is the assumption that $B(E_i,E_\mathrm{tr})$ is  independent of the initial 
state, that is, does not depend upon the initial energy $E_i$. 
The hypothesis is known to fail in systematic ways, especially as one goes up in the excited state energy spectrum~\cite{frazier1997gamow,PhysRevC.76.055803,MischBrink,SumRulesBrink}.  

As a test of the Brink-Axel hypothesis in calculations, one can consider moments or sum rules of the strength function,
specifically the non-energy-weighted sum rule or total strength, 
$S_0(E_i) = \int B(E_i, E_\mathrm{tr} ) \, dE_\mathrm{tr}$, and the energy-weighted sum rule,
$S_1(E_i) =\int E_\mathrm{tr} B(E_i, E_\mathrm{tr} ) \, dE_\mathrm{tr} $. Both are convenient to 
write as expectation values and so one can efficiently evaluate the sum rules for many states~\cite{SumRulesBrink,GenSumRules}. 
 If the Brink-Axel hypothesis were true, the sum rules would be independent of the initial energy 
 $E_i$.  For example, by looking at the first few transitions, one finds the centroids at 
 consistently the same location for both electric dipole~\cite{kruse2019no} and Gamow-Teller transitions~\cite{LMP}. 
Yet, by going up to systematically higher initial energies, recent work on several transition operators (electric quadrupole, magnetic dipole, and 
Gamow-Teller) has provided both numerical evidence and mathematical arguments that 
the sum rules are not and cannot be independent of energy.   However, it is crucial to note that the sum rules 
evolve smoothly with energy, exhibiting robust fluctuations, by which we mean that the fluctuations calculated in an 
energy bin are insensitive to the bin size 
\cite{SumRulesBrink,GenSumRules}. This observation leads to a modification of the Brink-Axel hypothesis, as we discuss in the next section.
 


\subsection{A brief history of calculating astrophysical weak rates, and the goal of this paper}

Where possible, measured values of Gamow-Teller transition strengths are used in calculations of 
astrophysical weak rates. But such experimental data comprise only a  fraction of the input needed, especially 
for hot, dense environments where nuclei inhabit highly excited initial states to a significant proportion. The rest must be supplied by theory.

The foundational work of Fuller, Fowler and Newman (FFN) \cite{FFNI,FFNII,FFN3,FFNIV} computed stellar weak rates for mass number $A=17-60$ on temperature and density 
grids  appropriate to pre-collapse massive stars. FFN used a variety of nuclear data, for example measured energy levels and measured $ft$-values for weak transitions, isopsin symmetry where exploitable, and extant $sd$-shell shell model calculations of wavefunctions and Gamow-Teller matrix elements. FFN employed an Independent Particle model to estimate remaining strength in, and the excitation energy centroid of, Gamow-Teller (GT) resonances. They  applied the Brink-Axel hypothesis to approximate the centroid excitation energies of Gamow-Teller resonances built on excited states.  

Later works used large scale configuration-interaction shell model calculation, relying upon increasing 
computational power, 
starting with the 1994 updates of weak transitions in $sd$-shell nuclei, $A=17-39$ by Oda et al. \cite{OdaEtAl}. By computing 
at least 100 levels in $sd$-shell parent nuclei,  they could capture virtually all of the relevant thermal weak transition strength. These rates were appropriate for for the oxygen, neon, and magnesium burning stars of intermediate mass below 10 solar masses. 
(There have been some recent updates \cite{PhysRevC.88.015806,PhysRevC.89.045806}.)
For massive stars of 10 solar masses and higher, $pf$-shell nuclides dominate, but the required model spaces are 
significantly larger and computationally much more challenging.

Following on the heels of a preliminary configuration-interaction study by Aufderheide \textit{et al.}~\cite{PhysRevC.48.1677}, 
the 1999 study of Caurier, Langanke,  Martinez-Pinedo {and Nowacki}~\cite{CLMP} (CLMPN) found that configuration-interaction calculations 
in the $pf$-shell, 
using the KB3 shell-model interaction of Poves and Zuker \cite{kb3} yield better placement of the Gamow-Teller centroid than the FFN Independent Particle Model. Furthermore, while the Independent Particle Model puts all the Gamow-Teller strength at one place in the final spectrum, configuration-interaction calculations correctly fragment the Gamow-Teller strength 
over many final states, a consequence of SU(4) symmetry breaking.

A year later  Langanke and Martinez-Pinedo (LMP) \cite{LMP} updated rates for the $pf$-shell nuclides with $A=45-65$. 
While LMP included heavy $pf$-shell nuclides beyond those in FFN, they only computed states 4 to 12 parent eigenstates,  limiting excitation energy to 1 MeV for odd-$A$ and odd-odd nuclides, and to 2 MeV for even-even nuclei. Following FFN, 
however, LMP supplemented these calculations with  the Brink-Axel Hypothesis.  Pruet and Fuller followed up 
by looking at nuclides with $A=65$-$80$, using experimental data, trends in Gamow-Teller systematics, and 
explicit shell model calculations~\cite{pruet2003estimates}. 

As an alternative, the quasi-particle random-phase approximation (QRPA) can 
compute Gamow-Teller transitions strengths in larger model spaces than those accessible by configuration-interaction calculations, especially when using a solvable schematic interaction~\cite{nabi2004microscopic,nabi2005gamow,PhysRevC.76.055803}.  Such schematic calculations, however, tend to 
have significantly different strength functions even from random phase approximation calculations using 
realistic shell-model forces \cite{nabi2013comparison}. Related thermal QRPA calculations using 
Skyrme interactions~\cite{PhysRevC.81.015804,PhysRevC.100.025801,PhysRevC.101.025805} can also reach highly excited transitions, but may underestimate the total rates due to 
missing correlations~\cite{PhysRevC.103.024326}. (In ground state proton-neutron RPA calculations, fragmentation of the 
Fermi surface due to breaking of axial symmetry can also increase Gamow-Teller strength; furthermore, at least in the lower $pf$-shell, breaking 
of rotational symmetries has a stronger effect than breaking particle number 
in spherical proton-neutron QRPA calculations~\cite{PhysRevC.69.024311}. However, it is not clear how this would be reflected 
in finite temperature calculations.)

Following earlier work on the systematics of Gamow-Teller strengths \cite{frazier1997gamow} and 
moments of strength functions \cite{SumRulesBrink},
Misch, Fuller and Brown  examined the validity of the Brink-Axel hypothesis in $sd$-shell nuclei, where with more powerful computers, they were easily able to obtain initial states of up to 28 MeV. They found the BAH fails at low and moderate initial excitation energy, but may have validity at high initial excitation energy~\cite{MischBrink}. 
Our work here can be considered  an extension of that earlier study.

Today's desktop computers are  able to obtain many hundreds of eigenpairs in the $sd$-shell, i.e., 
nuclides between $^{16}$O and $^{40}$Ca, after many thousands of Lanczos iterations, allowing one to directly compute transitions of interest even between 
highly excited states. 
But in the $pf$-shell, that is, nuclides between $^{40}$Ca and $^{80}$Zr, the problem is much more 
challenging. Although using supercomputers one can obtain low-lying states, direct access to 
highly excited states in $pf$-shell nuclides is generally still out of reach. 

In this work, we explore the regularities in Gamow-Teller strength functions \textit{\`a la} Brink-Axel for several transitions in the iron region:
\begin{equation}
\begin{aligned}
^{53}\mathrm{Fe} &\rightarrow\, ^{53}\mathrm{Mn} \, \\
^{55}\mathrm{Fe} &\rightarrow\, ^{55}\mathrm{Mn} \, \\
^{55}\mathrm{Cr} &\rightarrow\, ^{55}\mathrm{Mn} \, \\
^{56}\mathrm{Fe} &\rightarrow\, ^{56}\mathrm{Mn} \, \\
^{57}\mathrm{Co} &\rightarrow\, ^{57}\mathrm{Fe} \,.
\end{aligned}
\label{eqn:targets}
\end{equation}
While the Brink-Axel hypothesis does not strictly hold, a generalization does: strength functions 
from initial states nearby in energy behave similarly, a finding which tracks previous 
work on GT strength functions~\cite{MischBrink} and moments of strength functions~\cite{SumRulesBrink}. Exploiting this
 energy-localized Brink-Axel hypothesis,  we are able to obtain transitions strengths for highly excited, semi-converged states, which can be then directly applied in thermal Gamow-Teller (GT) transition rate computations in massive stellar cores.

From our strength functions for $^{57}\mathrm{Co} \rightarrow\, ^{57}\mathrm{Fe}$ 
 we  compute thermal weak rates at  temperatures and densities found in massive star core conditions preceding collapse. We find very good agreement with LMP thermal Gamow-Teller rates through most of the density and temperature range up to a temperature of $10^{10}$ K.  Beyond this temperature, there is a sudden uptick in the rates and their growth that could have significant astrophysical implications.  
 More broadly, however, our modification to the Brink-Axel hypothesis opens a path to more microscopic 
 foundations for computing weak rates in extreme environments. 

\section{Methods: Calculation of weak transitions using the nuclear shell model}

\label{SM}

We work in the interacting shell model~\cite{BG77,ca05},  a subset of configuration-interaction methods.  The many-body Schr\"odinger equation is written as a matrix 
eigenvalue problem by expanding  in a convenient orthonormal basis,
$| \Psi \rangle = \sum_\alpha v_\alpha | \alpha \rangle $.  Here the basis states $\{ | \alpha \rangle \}$ are antisymmeterized products of 
single-particle wave functions, usually written in second quantization, that is, the occupation 
representation of Slater determinants. The single-particle basis itself is organized into orbitals defined by good angular momentum quantum numbers.   
The interacting shell model has the advantage of handling even and odd numbers of particles equally well. Most importantly, one can generate many excited states, and in principle can 
obtain every eigenstate in the model space; this contrasts with methods 
primarily designed to 
find the ground state. 

The downside of the interacting shell model is the exponential explosion of 
 the basis dimensionality.  Calculations in the the $sd$-valence space, 
that is, nuclei with a $^{16}$O core and valence particles restricted to the $1s_1/2$-$d_{3/2}$-$0d_{5/2}$
space have a maximum basis dimension of 93,000 in the $M$-scheme (that is, fixed $z$-component of angular momentum or $M$);
but calculations in the $pf$-shell, that is, nuclei with a $^{40}$Ca core and 
$1p_{1/2}$-$1p_{3/2}$-$0f_{5/2}$-$0f_{7/2}$ valence space, have $M$-scheme basis  dimensions up to 2 billion.

Shell model codes routinely rely upon the Lanczos algorithm \cite{Whitehead} or a related variant 
to efficiently obtain low-lying states. The Lanczos algorithm constructs 
Krylov subspaces by repeated application of the Hamiltonian matrix on vectors, 
and approximate Hamiltonians are diagonalized within succeeding subspaces. As the subspace 
increases with each iteration, the extremal eigenpairs of the truncated 
Hamiltonian, the most relevant here being the low-lying eigenpairs, converge to those of the full Hamiltonian. 
While low-lying states of all the $pf$-shell nuclides are accessible, at least with the help of supercomputers, obtaining 
highly excited states remains a significant challenge. In the Appendix  we give more 
details of our use of Lanczos, including a modified Lanczos algorithm to sample highly excited states. 
To carry out our calculations we use the {\sc Bigstick} code \cite{BIGSTICK1,BIGSTICK2}, which has 
full Message Passing Interface (MPI) and OpenMP parallelization, making it suitable for applications on large supercomputers.

 We use the semi-empirical $pf$-shell  interaction GXPF1A \cite{honma2005shell}, which has smaller and more uniform deviations in lowing lying excitation energies compared to the classic KB3 family interactions \cite{kb3} \cite{kb3g}. The GXPF1 family of $pf$-shell residual interactions have 
also been shown to follow more closely experimental Gamow-Teller distributions in terms of the location of resonance peaks and fragmentation of the distribution \cite{PhysRevC.65.061301}. For the transition $^{57}\mathrm{Co} \rightarrow\, ^{57}\mathrm{Fe}$  we use initial excited  states going up to 80 MeV, requiring  the significant memory and processing capability of supercomputers. 
Compared to LMP we include  many more initial states, up to 100 to 200 converged states, 
roughly up to about 5-8 MeV in excitation energy. 
As in Oda et. al.~\cite{OdaEtAl} for most temperatures and densities this suffices to converge most of the thermal Gamow-Teller strength. But for temperatures above $10^{10}$ K even more excited states 
are needed. For those highly excited states we utilized our energy-localized Brink-Axel hypothesis, 
described in more detail below.

\subsection{Computing strength functions}

\label{strength_functions}

In this paper we consider three methods to generate strength functions. 
The first  is to painstakingly compute transition strengths between individual eigenstates. 
 Any one-body transition operator (such as the Gamow-Teller operator) with angular momentum rank $K$ and $z$-projection $M$ one can be expanded
\begin{equation}
\hat{\cal O}_{KM} = \sum_{a,b} O_{ab} \frac{1}{\sqrt{2K+1}}\left [ \hat{c}_a^\dagger \otimes \tilde{c}_b \right ]_{KM}   
\end{equation} 
where $\hat{c}_a^\dagger$ represents a creation operator for a particle in a spherical single-particle state, that is a 
state with good orbital and total angular momentum, labeled by $a$, and $\tilde{c}_b$ is a (time-reversed) annihilation operators for a state labeled by $b$; the square brackets denote coupling up to a good total angular momentum $K$ 
with $z$-component $M$. Finally, $O_{ab} = \langle a || \hat{\cal O}_K || b \rangle$ is the reduced matrix element 
between single-particle states.  This is useful because from the reduced one-body density matrix  between an initial state $\psi_i$ and a final state $\psi_f$,
\begin{equation}
\rho^{fi}_{K}(ab)=\sqrt{\frac{1}{(2K+1)}}\sum_{ab}\langle\psi_f||[\Hat{c}_a^\dagger\otimes\Tilde{c}_b]_{K}||\psi_i \rangle,
\end{equation}
 the reduced transition strength is simply
\begin{equation}
\langle\psi_f|| \hat{\cal O}_{K}||\psi_i \rangle = \sum_{ab}\rho^{fi}_{K,T}(ab) O_{ab}
\end{equation}
The $B$-value is given by Eq.~(\ref{eqn:BGT}). 
Note that this assumes that the states $i$ and $f$ have good angular momentum and are 
numerically exact eigenstates, i.e., they are fully converged in the Lanczos algorithm.  While this 
method is ironclad, 
because it requires converged eigenstates it can only be used between relatively 
low-lying states. Nonetheless, it is a useful check of other methods.

The second method, widely used, starts with a converged initial state and then efficiently
construct a strength function to  final states using a modified Lanczos algorithm~\cite{ca05,whitehead1980shell,bloom1984gamow}. This Lanczos technique was used to study Gamow-Teller strength distributions for some iron peak isotopes of astrophysical interest \cite{BloomFuller85}. 
Because this method assumes converged initial states, it is still limited to initial states of relatively low excitation energy when the basis dimension of the initial nuclide is especially large;  in the $pf$-shell eigenstates up to 5-10 MeV excitation energy may be possible but require thousands of Lanczos iterations. Once an initial state is obtained, however, it takes relatively few iterations to converge a good distribution of strengths to  final states.  For example, LMP used 33 iterations to generate strength functions; in our calculations we took 100 iterations.  While not all of the final states will be converged (the extremal ones will be), this approach nonetheless
obtains the moments of the strength function to a very high order~\cite{whitehead1980moment}, which can be and have been 
 checked against the first method. This application of Lanczos for strength functions has long been a workhorse for computing 
Gamow-Teller transitions for astrophysical applications.

Finally, in a new approach we carry out calculations 
using semi-converged initial states (generated using a different variant of Lanczos, also described in the Appendix) at much higher 
initial energy, and then compute the strength 
function following the Lanczos moments methods. This approach works due to the energy-localized 
Brink-Axel hypothesis: because  initial states nearby in energy have similar strength functions, 
even an initial state that is a linear combination of nearby states (i.e., has a small if nonzero 
dispersion in energy) can serve as a useful proxy for fully converged states. (These ideas were 
partially anticipated by Langanke and Martinez Pinedo~\cite{LMP}, 
who also included a few semi-converged `averaged GT states' at moderate excitation energy.  They argued that at high temperature and 
densities, variations in low-lying strength would tend to cancel, leading to a recovery of 
the Brink-Axel hypothesis. Our work here is a more in-depth and more systematic investigation of these ideas.) 

As a small technical point: in {\sc Bigstick} all calculations must be carried out in a basis with fixed $M$ and $M_T$.
Charge-changing transitions are computed first as charge-conserving (but isospin changing) transitions, always in the 
basis of smallest $M_T$. For example, for $^{57}$Co ($M_T= 7/2$) $\rightarrow ^{57}$Fe ($M_T=5/2$), we work in the 
basis for $^{57}$Fe but compute transitions from and to the isobar analogs of states of $^{57}$Co. We can access these 
as initial states by adding $-\lambda T(T+1)$ to the Hamiltonian to bring them down in energy. 
Afterwards we 
use isospin Clebsch-Gordan coefficients to obtain the  charge-changing matrix elements.

\begin{figure}
    \centering
    \includegraphics[scale=0.5,clip]{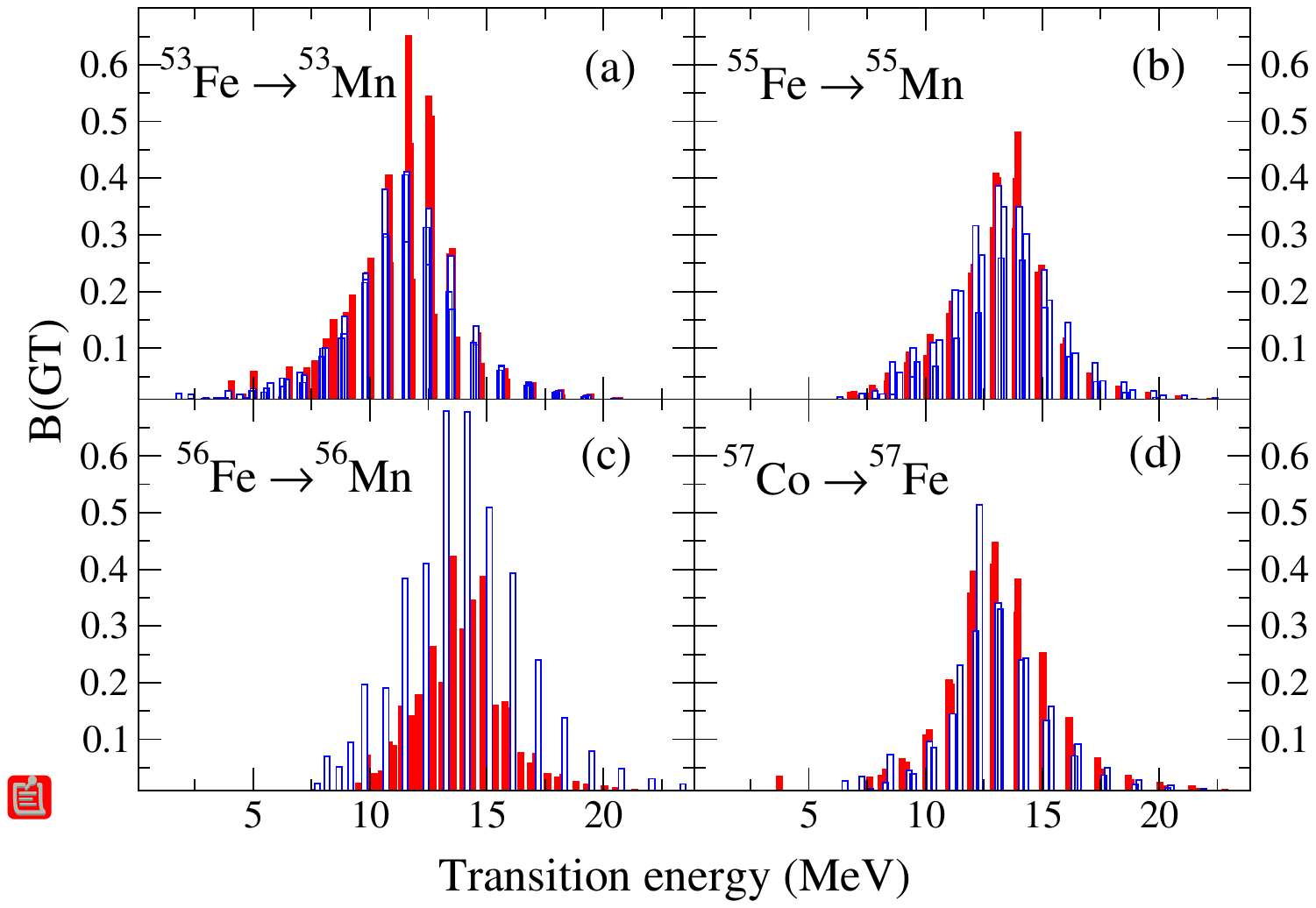}
    \caption{Strength functions from two adjacent initial  states (filled and open bars) with excitation energies near 4.5 MeV, 
for (a) $^{53}$Fe$\rightarrow ^{53}$Mn;  (b) $^{55}$Fe$\rightarrow ^{55}$Mn;  (c) $^{56}$Fe$\rightarrow ^{56}$Mn;  
and  (d) $^{57}$Co$\rightarrow ^{57}$Fe.} 
    \label{fig:BrinkAll}
\end{figure}

\section{Results:  Gamow-Teller Transition Strengths and the Brink-Axel hypothesis} \label{sec:GTandBA}

In order to compute thermal rates in stellar environments, we need the reduced transition probabilities, $B(GT)$, defined in Eq.~(\ref{eqn:BGT}),
or the related strength function, Eq.~(\ref{eqn:strfunc}), which depend upon both the initial state and the transition energy, $E_\mathrm{tr} = E_f-E_i$.  When possible one uses the experimental final  and initial energies, $E_f$ and $E_i$, respectively, but these are often not available. In that case one adopts 
\begin{equation}
    E_\mathrm{tr}=Q_0+E^\mathrm{SM}_f-E^\mathrm{SM}_i,
    \label{eqn:cue}
\end{equation}
where $E^\mathrm{SM}_f,E^\mathrm{SM}_i$ refer to the excitation energies as computed in the configuration-interaction shell model, 
and $Q_0=m_{re}-m_{pr}$ is the Q-value,  the experimental difference in mass of the reactants and products in their ground states.
Note that in our strength function plots we use the shell-model Q-value, although for the thermal rates, we use the experimental $Q_0$; the difference is just an overall shift in energy.

The central idea in this paper is an energy-localized Brink-Axel hypothesis, namely that the Gamow-Teller strength function from states nearby in 
energy are similar. An illustration of this is in  Fig.~\ref{fig:BrinkAll}, showing  four of the five  transitions
listed in Eq.~(\ref{eqn:targets}). For each set of nuclides we show the strength function 
from two nearby initial states at around 4.5  MeV 
in excitation energy. 

While one can see the strength functions look similar, the fragmentation of strength can make comparison 
difficult.  One strategy for comparison is to fold strengths with an appropriate resolution function~\cite{PhysRevC.72.065501}.
We instead use the running sum of the strength function: 
\begin{equation}
    R(GT)[E_i,E_f]=  \sum_{f^{\prime} \leq f } B_{if^\prime}(GT),
    \label{eqn:RGT}
\end{equation}
As shown in 
Fig.~\ref{fig:strfunc} in the Appendix, the running sum quickly converges as the number of Lanczos 
iterations increases, making for reliable comparisons.


\begin{figure} 
    \centering
    \begin{tabular}{cc}
    \includegraphics[scale=0.325,clip]{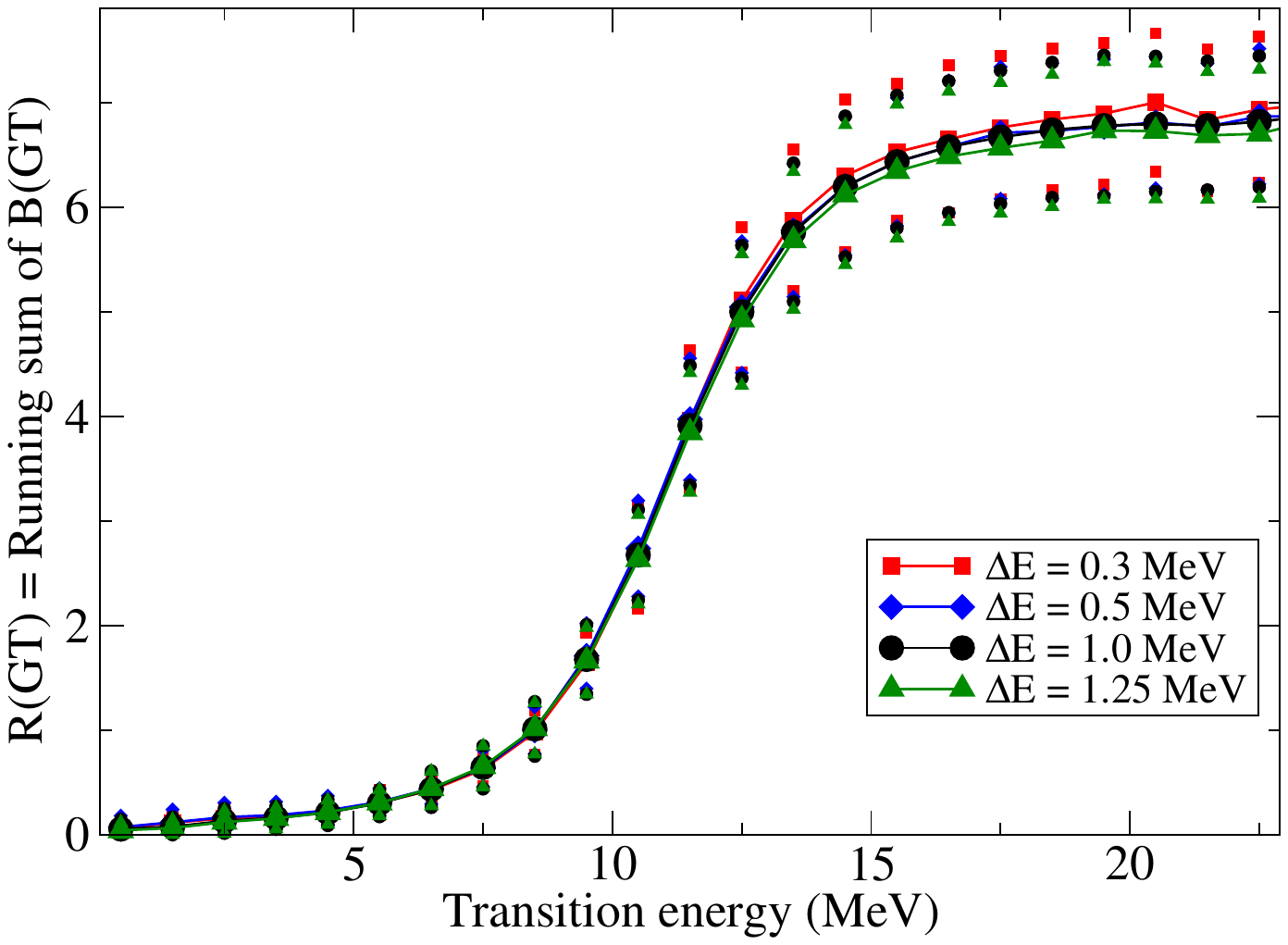}
            \includegraphics[scale=0.325,clip]{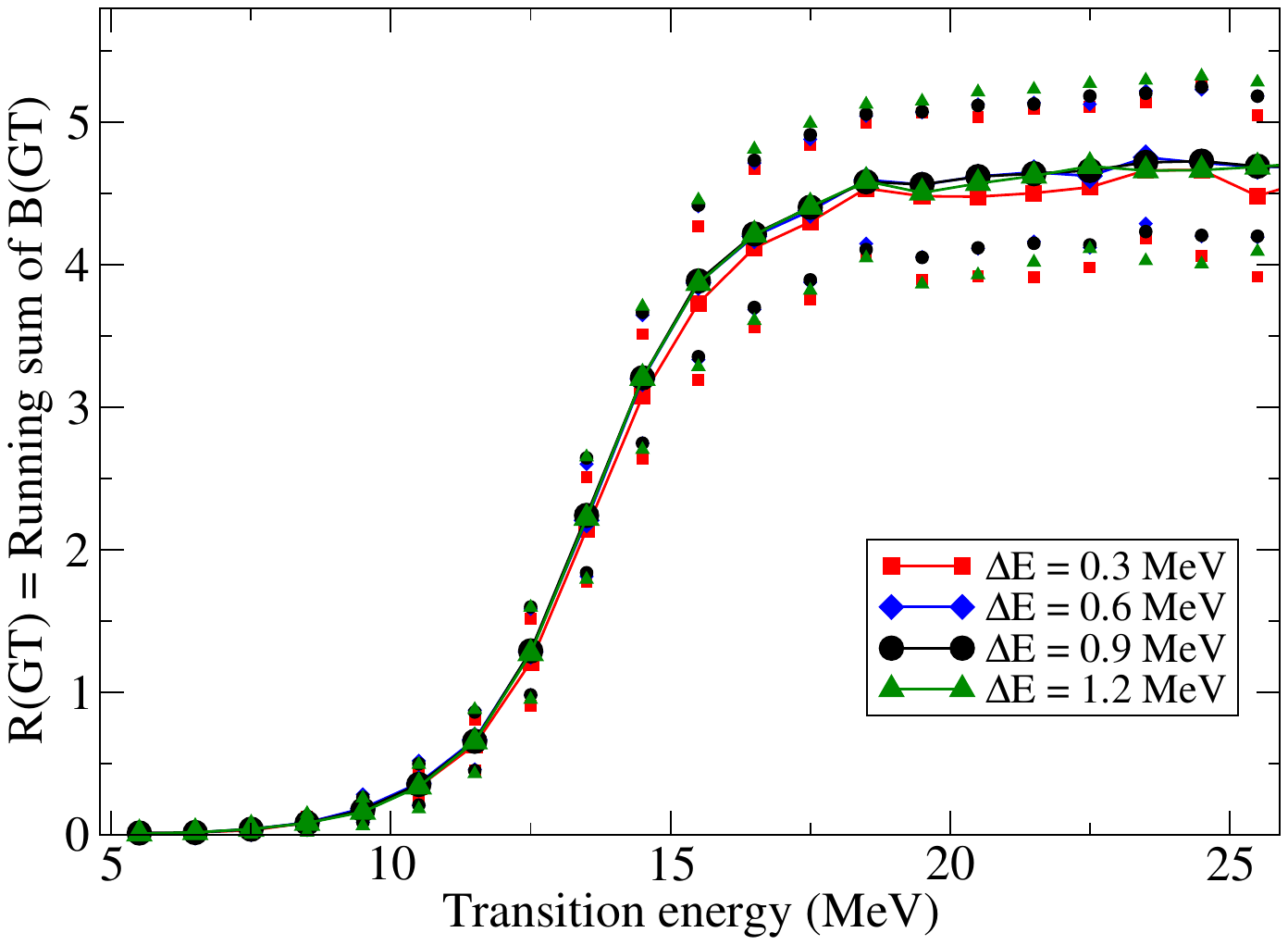}
    \end{tabular}
    \caption{Running sums, Eq.~(\ref{eqn:RGT}), for  (a) $^{53}\mathrm{Fe} \rightarrow\, ^{53}\mathrm{Mn}$ for converged initial states around 3.96 MeV. 
Data is in initial energy bins of size from 0.3 to 1.25 MeV; shown are the average (large connected symbols) and standard deviation (small, 
unconnected symbols); and (b)   $^{56}\mathrm{Fe} \rightarrow\, ^{56}\mathrm{Mn}$ with  converged initial states around 4.25 MeV, 
in energy bins from 0.3 to 1.25 MeV.}
    \label{fig:Brink1}
\end{figure}


As discussed in Section \ref{BAHbasics},
the Brink-Axel hypothesis does not hold rigorously for Gamow-Teller transitions.  Yet work on summed strengths 
\cite{SumRulesBrink} provides powerful evidence for a secular evolution of strength functions. In that work 
the total Gamow-Teller strength, which is just the integral of the strength function, was calculated for 
every initial state; this was done efficiently as an expectation value. The summed strengths were then put into bins of initial energy and the average and the 
width (standard deviation) in each bin calculated. The widths were largely insensitive to the size of the energy bins, and both the 
average and the widths evolved smoothly with energy. 

These observations instruct our work here and lead us to suggest the energy-localized Brink-Axel hypothesis, 
namely that \textit{strengths from initial states nearby in energy are similar, within well-defined statistical fluctuations}. This was already observed in full calculations  
of Gamow-Teller strength functions in the $sd$-shell  \cite{MischBrink}. In the rest of this paper we 
provide further evidence for this ELBAH in $pf$-shell nuclides. We also invoke the ELBAH for a practical 
calculation of strength functions from highly excited states: if strength functions from nearby initial eigenstates of the initial Hamiltonian  
are similar, we don't need fully converged states. Instead, we only need semi-converged initial states 
that are a superposition of states nearby in energy. 

Following \cite{SumRulesBrink}, we compute the statistics of the running sums $R(GT)$ as follows:
Given a chosen initial energy $\bar{E}_i$, we take running sums with initial energies in a bin of size $\Delta E$, that is, 
$\bar{E}_i -\Delta E/2 \leq E_i \leq \bar{E}_i + \Delta _i$. We then take the running sums to final energies $E_f$ 
in \textit{fixed} bins of width 1.0 MeV, and compute the averages and widths (standard deviation). We then plot 
the averages and widths for several different initial energy bin sizes $\Delta E$. 
Our results, as exemplified in following figures, show that the averages and the widths are largely 
insensitive to the size of the initial energy bin, meaning the running sums (and thus the original strength functions) are  similar within fluctuations; if the strength functions were not similar, we would expect a sensitive dependence upon the size of the initial
energy bin.


\subsection{Converged initial states: the standard approach}

Up until now, even in calculations~\cite{LMP} using the Lanczos strength function `trick' \cite{ca05,whitehead1980shell,bloom1984gamow} (the second of the 
three approaches discussed in Section \ref{strength_functions}), one almost always used initial states that had converged under the original Lanczos algorithm; by this we mean 
eigenstates of the nuclear Hamiltonian found within the numerical accuracy 
of the code. (The exception is LMP who used some 
`averaged' states~\cite{LMP},  presaging the work discussed in the next section.) 
Here we give two examples with converged initial states in Fig.~\ref{fig:Brink1} for the $^{53}\mathrm{Fe} \rightarrow\, ^{53}\mathrm{Mn}$ transition centered at 3.96 MeV, 
and 
$^{56}\mathrm{Fe} \rightarrow\, ^{56}\mathrm{Mn}$ transition centered at 4.25 MeV. 
In both cases the total Gamow-Teller strength (using $g_A = 1$) have a variance of around 0.5-0.6. A peak occurs  
at the inflection point, around 11 MeV for Fig.~\ref{fig:Brink1}(a) and 14 MeV for Fig.~\ref{fig:Brink1}(b).  

The problem with this approach is the computational burden to reach higher energies. For example, for $^{49}$Cr, which has 
an $M$-scheme dimension of only 6 million basis states, the first 100 levels only gets one to 5.5 MeV in excitation energy, 
and the first 500 levels, which requires about 7000 iterations using thick-restart Lanczos~\cite{wu2000thick}, only gets one to 8.28 MeV in excitation energy. While for many astrophysical calculation these initial excitation energies may be sufficient, 
they also require heroic computational efforts. Fortunately the energy-localized Brink-Axel hypothesis suggests a way forward.




\subsection{Semi-converged initial states} \label{sec:semi-con}

As we just discussed, for even modest excitation energy it can take many Lanczos iterations to converge states (and in general the problem of 
finding interior eigenpairs is notoriously difficult). This is arguably because of the 
high level densities, or conversely the small energy differences between levels. Yet the high level density, hand-in-hand with the 
ELBAH, leads to an alternate, less demanding approach. Since ELBAH tells us that states nearby in energy have similar strength functions, 
we do not need fully converged initial states. Instead we can use semi-converged states, that is, states that are linear combinations 
of eigenstates (corresponding to fully converged states) nearby in energy. As the strength functions from nearby 
states are statistically similar, we can use admixtures of nearby states as proxies for the `true' eigenstates. 

Thus we introduce a third approach for strength functions outlined in Section \ref{strength_functions}. It is only necessary to have semi-converged states with a relatively 
modest variance in the energy $\langle \hat{H}^2\rangle-\langle \hat{H}\rangle^2$, around (0.5 MeV)$^2$ to (1.0 MeV)$^2$.  
These states we find by a variant of thick-restart Lanczos, which we term \textit{targeted thick-restart Lanczos} and which we describe 
in the Appendix.  These semi-converged states will not have good quantum numbers $J$ (angular momentum) or $T$ (isospin) but we 
can use, again, the Lanczos algorithm to project out components with good quantum numbers~\cite{morrison1974novel,PhysRevC.91.034313,PhysRevC.95.024303}. 

\begin{figure}
    \centering
    \includegraphics[scale = 0.5,clip]{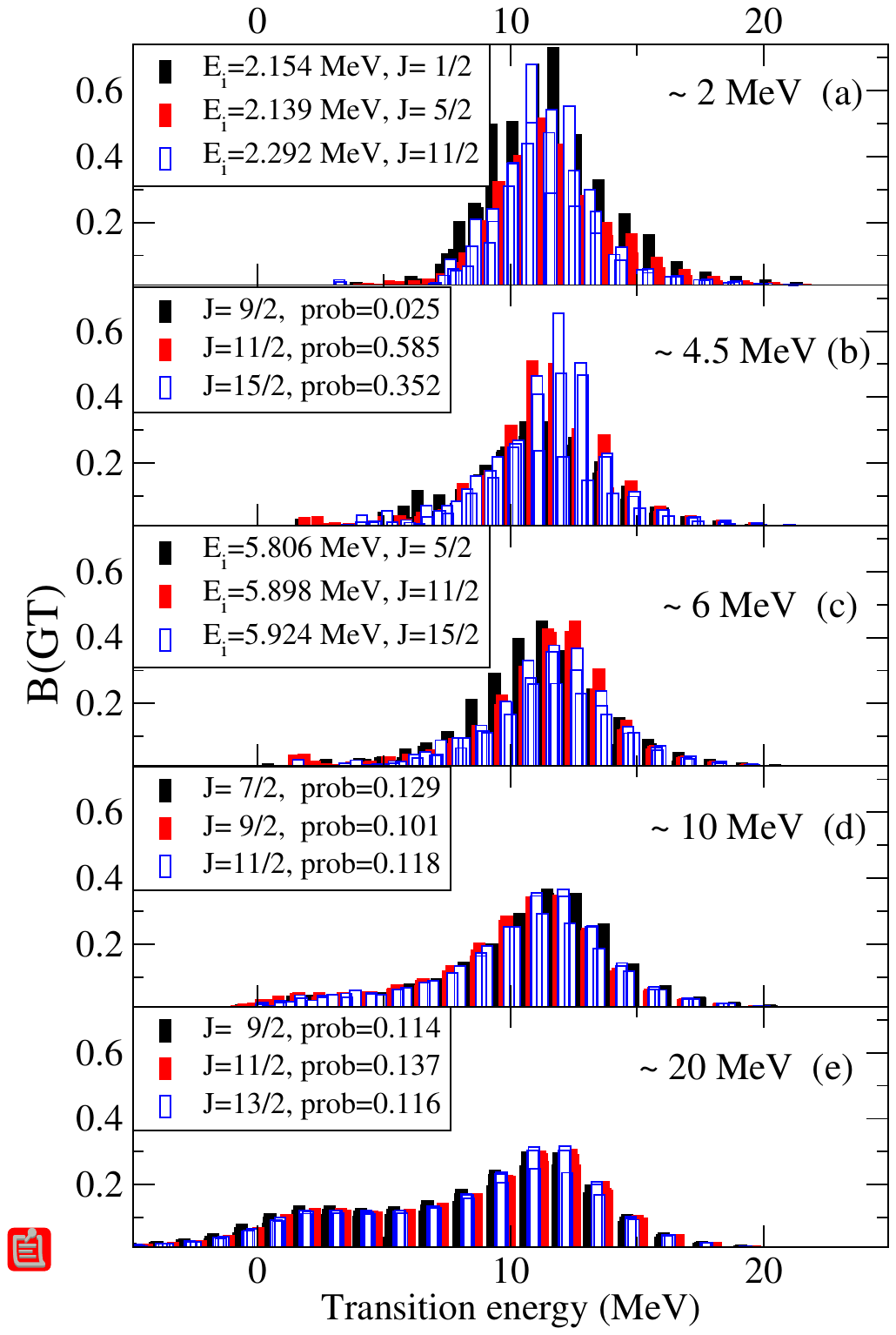}
    \caption{Evolution of the strength function for $^{53}\mathrm{Fe} \rightarrow\, ^{53}\mathrm{Mn}$ as one goes higher in initial excitation energy.  Panels (a) and (c) are for selected fully converged states, with initial energies and angular momenta $J$ given.
Panels (b), (d), and (e) are for semi-converged initial states around energies 4.5, 10, and 20 MeV, respectively. The probabilities 
are the fraction of the semi-converged state; not all components of different $J$ are shown, hence the probabilities here do not sum to 1.}
    \label{fig:FeMn53evol}
\end{figure}


In Fig.~\ref{fig:FeMn53evol} we plot strength functions (not running sums) from initial states nearby 
in energy at several different energy regimes.  We selected fully converged states from around 2 MeV in initial energy in Fig.~\ref{fig:FeMn53evol}(a), and  from around 
6 MeV in Fig.~\ref{fig:FeMn53evol}(c): we give both the 
initial energies and the initial angular momenta $J$.  We used semi-converged states from around 4.5 MeV in Fig.~\ref{fig:FeMn53evol}(b), 10 MeV in Fig.~\ref{fig:FeMn53evol}(d), 
and 20 MeV in Fig.~\ref{fig:FeMn53evol}(e),  and projected out states of good $J$ and isospin $T$.
We give $J$ and the fraction (probability) these states are of the original semi-converged states; as not all $J$ values 
are shown, the probabilities do not sum to 1. 

What should be apparent is first, there is a steady overall evolution of the strength function, with, second, different 
initial states having similar albeit not identical strength functions and, third, no obvious difference in the behavior 
between the strength function plots of converged and semi-converged initial states.

\begin{figure}
    \centering
    \begin{tabular}{cc}
    \includegraphics[scale=0.33,clip]{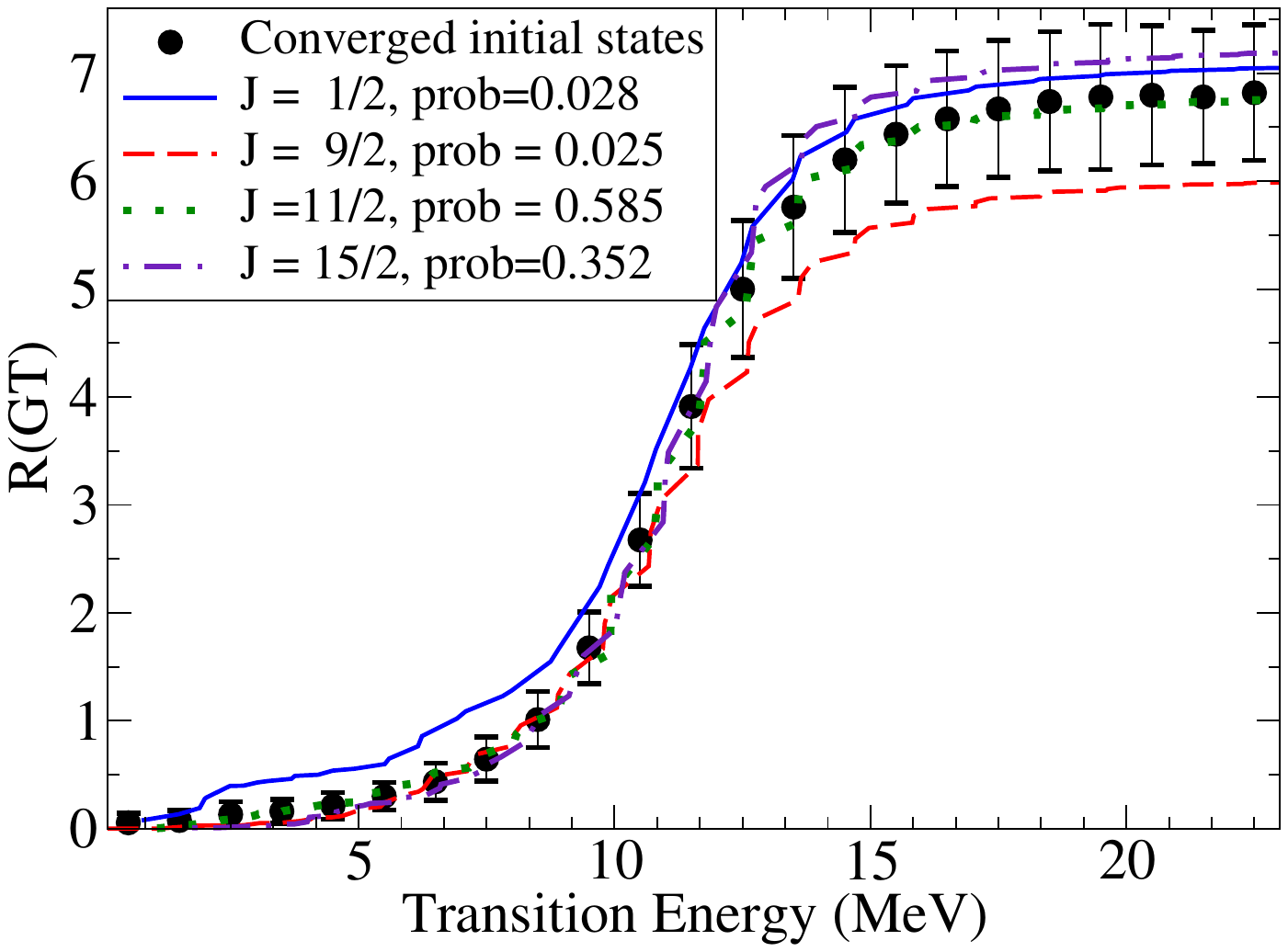}
        \includegraphics[scale=0.33,clip]{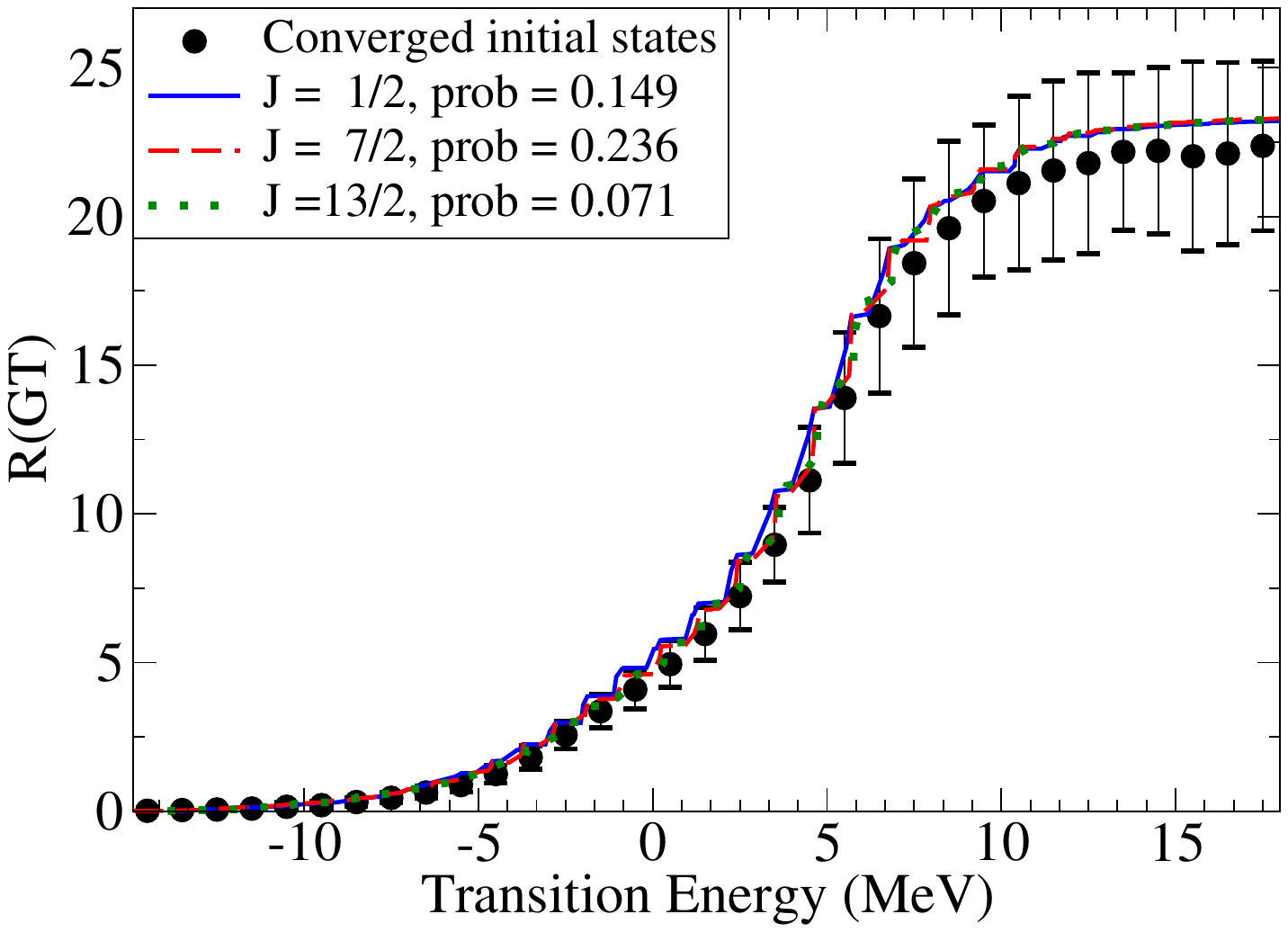}
    \end{tabular}
    \caption{(a) Comparison of running sums, Eq.~(\ref{eqn:RGT}), for converged states average around 4.5 MeV excitation (filled circles with error bars) and individual semi-converged states also around 4.5 MeV excitation (colored, connected lines) in the $^{53}\mathrm{Fe} \rightarrow\, ^{53}\mathrm{Mn}$ transition.
(b) Comparison of running sums for converged states average around 4 MeV excitation (fill circles
with error bars) and individual semi-converged states around 5 MeV excitation (colored, connected lines) in the $^{55}\mathrm{Cr} \rightarrow\, ^{55}\mathrm{Mn}$ transition}
    \label{fig:FeMn53_4.5}
\end{figure}

The running sums, $R(GT)$, yield similar conclusions. Fig.~\ref{fig:FeMn53_4.5} compares running sums of strengths 
functions from both binned averages and widths of $R(GT)$ from converged states, and $R(GT)$ from individual semi-converged initial states around 4-5 MeV, 
for $^{53}$Fe $\rightarrow ^{53}$Mn in Fig.~\ref{fig:FeMn53_4.5}(a), and  for $^{55}$Cr $\rightarrow ^{55}$Mn in Fig.~\ref{fig:FeMn53_4.5}(b).  The running sums from the semi-converged states broadly follow the statistical behavior 
of the converged initial states.


\begin{figure}
    \centering
    \begin{tabular}{cc}
    \includegraphics[scale=0.33,clip]{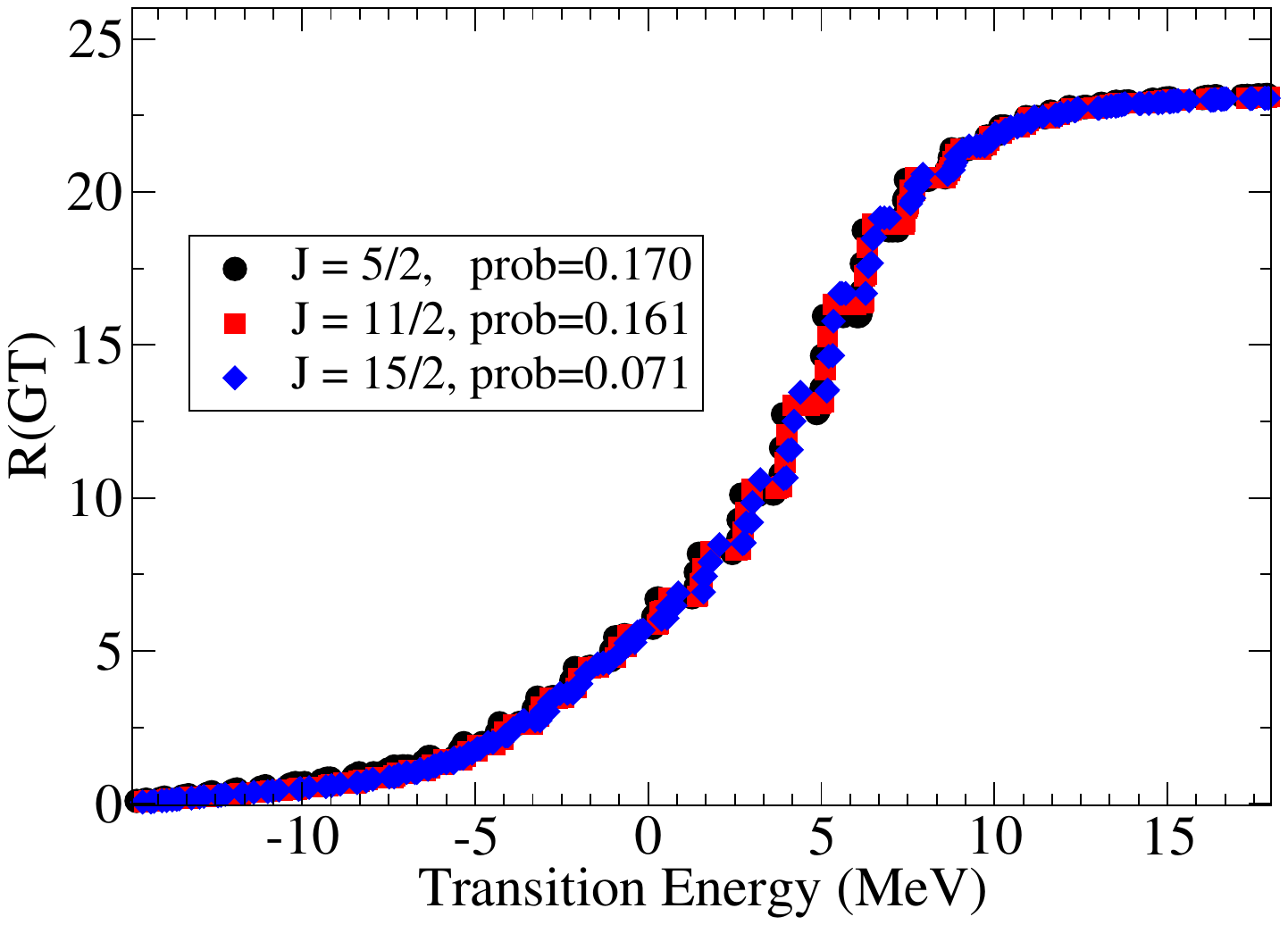}
        \includegraphics[scale=0.33,clip]{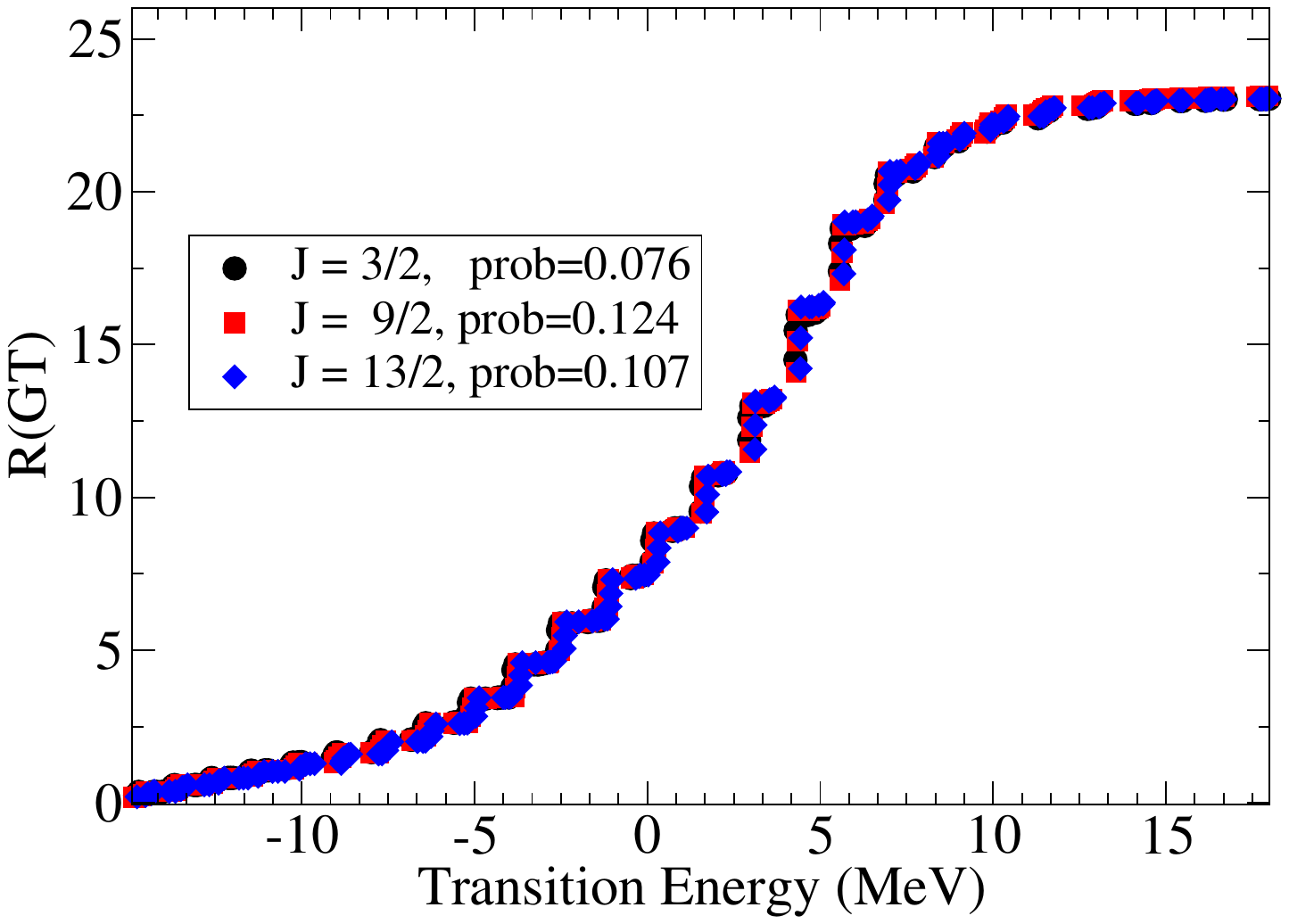}
    \end{tabular}
    \caption{Transition strength running sums, Eq.~(\ref{eqn:RGT}), for $^{55}\mathrm{Cr} \rightarrow\, ^{55}\mathrm{Mn}$ for semi-converged initial states around (a) 10 MeV and (b) 20 MeV}
    \label{fig:CrMn55_10_20}
\end{figure}

Because of computational constraints one cannot generally go much beyond 5 to 8 MeV in initial energy if fully converged states are used. Thus for higher excitation energies we  use only projected semi-converged states; 
we give examples from $^{55}$Cr$\rightarrow ^{55}$Mn, from initial states around 10 MeV in Fig.~\ref{fig:CrMn55_10_20}(a) 
and  from initial states around 20 MeV in Fig.~\ref{fig:CrMn55_10_20}(b). 
At these higher energies the running sums become nearly identical, echoing an earlier finding 
in the $sd$ shell \cite{MischBrink}. 
We did note a slight but consistent trend of initial states with higher $J$ 
having smaller total strength, though only at the few percent level.

\begin{figure}
    \centering
    \includegraphics[scale=0.33,clip]{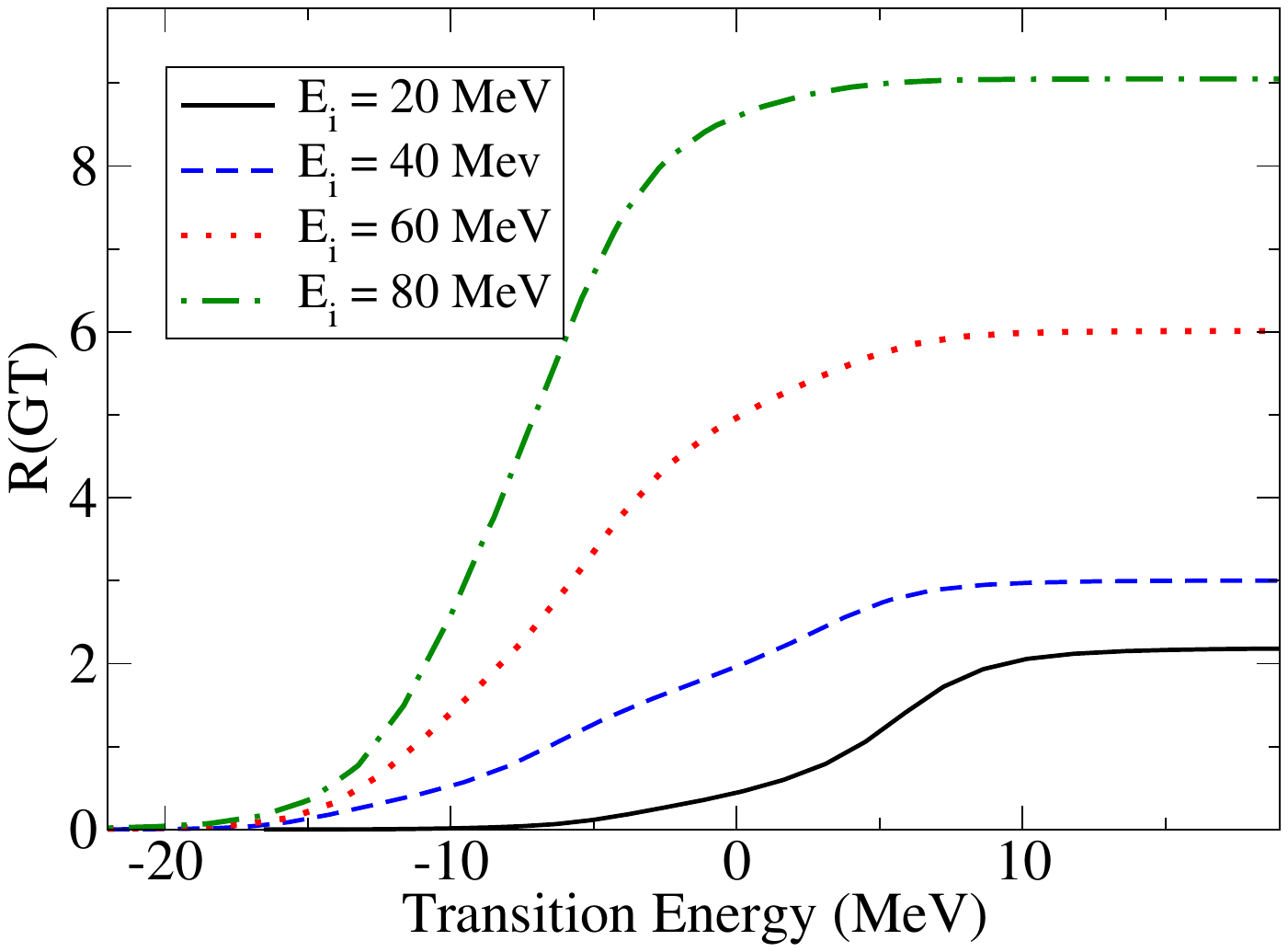}

    \caption{Transition strength running sums, Eq.~(\ref{eqn:RGT}), for $^{57}\mathrm{Co} \rightarrow\, ^{57}\mathrm{Fe}$ for semi-converged initial states at high initial excitation energy $E_i$.}
    \label{fig:CoFe57highEi}
\end{figure}

Finally, as an example of the power and applicability of this method, in Fig.~\ref{fig:CoFe57highEi}, we give sample strength function distributions 
from $^{57}\mathrm{Co}\rightarrow ^{57}\mathrm{Fe}$ from initial semi-converged states 
at around 20, 40, 60, and 80 MeV in initial excitation energy.  These strength function are used 
in the next section. One can see a clear evolution in the strength function, both in total strength 
and in the location of the centroid.  We note, however, as discussed in Section~\ref{background}, the average 
excitation energy for $^{57}$Co in this space is only 63 MeV; the lowering of the strength function centroids 
as one goes up in initial excitation energy is a necessary consequence of this. The fact that the total strength 
also evolves, however, cannot be explained by a finite model space.

Thus we have strong empirical evidence both for the ELBAH as well as the effectiveness of its application.

\section{Application to Massive Stellar Thermal Rates}

\label{application}

In astrophysical applications, such as computing the composition of pre-collapse massive stars, temperatures in billions of degrees Kelvin mean higher energy initial states can contribute significantly to thermal rates. 
Of relevance here, weak allowed rates 
determine the lepton to baryon ratio and the neutrino flux,  vital inputs to the final collapse dynamics of a star and its  products.

The partial transition rate from initial level $i$ to final level $f$,
$\lambda_{if}= B_{if} \Phi_{if}(Q_{if}, \rho, T),$ 
depends upon the transition strength $B_{if}$, which has been the focus of most of this paper, and the phase-space factor $\Phi_{if}$. 
The phase space factor depends not only the $Q$-value but also upon the process, such as  electron capture (we assume no energy or momentum dependence in the matrix element, consistent with an {\it allowed} transition, specifically Gamow-Teller or Fermi in the cases considered here) : 
\begin{equation}
    \Phi_{ij}^{(ec)}(\mu_e,T)=\int_{W_{min}}^{\infty} W\sqrt{W^2-1}(Q_{ij}+W)F(Z,w)f(T,\mu_e,+\epsilon)dW,
\end{equation}
or $\beta^+$-decay (positron emission),    
\begin{equation}
    \Phi_{ij}^{(pd)}(\mu_e,T)=\int_{1}^{W_{max}} W\sqrt{W^2-1}(Q_{ij}-W)F(-(Z-1),w)f(T,\mu_e,-\epsilon)dW,
\end{equation}
with $W_{max}=Q_{ij}$; and $W_{min}=1$ (in units where $m_e c^2 =1 $), if $Q_{ij}>-1$, or else $W_{min}=|Q_{ij}|$. As usual $F(Z,W)$ is  the Fermi function accounting for the attraction (repulsion) of the electron (positron) and the nucleus:
\begin{equation}
    F(Z,w)=2(1+\gamma)(2pR)^{-2(1-\gamma)}\frac{|\Gamma(\gamma+i\alpha Zw/p)|^2}{|\Gamma(2\gamma+1)|^2}
\end{equation}
with $\gamma=\sqrt{1-(\alpha Z)^2}$, where $\alpha$ is the fine structure constant, and $R$ is the nuclear radius. Finally $f(T,\mu_e;W m_e c^2)$ are the Fermi-Dirac statistics for the electron/positron:
\begin{equation}
    f(T,\mu_e,\epsilon)=\frac{1}{\exp{\left(\frac{\epsilon-\mu_e}{kT}\right)+1}}
\end{equation}
Here $\mu_e$ is the electron chemical potential and $p=\sqrt{W^2-1}$ is the momentum, again in units of the electron mass, 
while the total energy of the electron, including the rest-mass, is $\epsilon=W m_e c^2$. We assume the positron chemical 
potential is zero.  For the transition energy in 
Eq.~(\ref{eqn:cue}), $Q_0=M_p-M_d$ in units of $m_ec^2$ with $M_p$ and $M_d$ the  masses of the initial and final nuclei, respectively. The chemical potential was derived from the integral equation
\begin{equation}
    \rho Y_e = \frac{(m_e c)^3}{\pi^3\hbar^3 N_A}\int_1^\infty W\sqrt{W^2-1}[f(T,\mu_e,\epsilon)-f(T,-\mu_e,\epsilon)]dW,
\end{equation}
where  $N_A$ is Avogadro's number.

Phase space factors account for the final states of the electron/positron and the neutrino, where the Q-value, $Q_{ij}$, goes to their mass and kinetic energy depending on the reaction type. As we have demonstrated, 
within initial energy windows from 0.3-1.3 MeV strength functions have similar behavior, with 
decreasing fluctuations as one goes up in initial energy. This is particularly true and useful for initial state excitations above 5 MeV, where the capability to obtain converged eigenstates is limited. In this region we exploit the ELBAH and 
assume that within 2 MeV energy windows 
the reduced transition probabilities, $B(GT)$ 
are identical. 
Because at high excitation energy we do not explicitly include all transitions, but use the strength functions from 
sampled, semi-converged states as a proxy for the average strength function, we must include the number of levels 
in any energy window.
Therefore the level density plays a significant role in these calculations. 


\subsection{Case study: $^{57}\mathrm{Co} \rightarrow\, ^{57}\mathrm{Fe}$}

\label{casestudy}

As an application of our new approach, we calculated the thermal Gamow-Teller rates for electron capture and positron emission in $^{57}\mathrm{Co} \rightarrow\, ^{57}\mathrm{Fe}$, a case with  nearly 1 billion $M$-scheme basis states; not only is this system  a good test of the {\sc Bigstick} code \cite{BIGSTICK2}, but it is also expected to be important in  pre-collapse stellar evolution, {in particular in its contribution to the pre-collapse neutrino luminosity}~\cite{patton2017neutrinos}.  We compare the rates derived from our strength functions to  those of Fuller, Fowler, and Newman (FFN) \cite{FFN3}  and Langanke and Martinez-Pinedo (LMP) \cite{LMP}, as tabulated in \cite{sullivan2015sensitivity}, using the same methodology for computing rates as described in the introduction to this section.  Following the practice of those and most other authors, we used experimental 
data where it exists~\cite{bhat1998nuclear}, specifically: the experimental Q-value; the 
measured $B(GT)$ values (derived from the $\log f t_{1/2}$ values) from the 
$7/2^-$ ground state of $^{57}$Co to the first two excited $7/2^-$  states of $^{57}$Fe, which has a $1/2^-$ ground state; and the experimental excitation energies of those states.

\begin{figure}
\begin{tabular}{cc}
    \includegraphics[scale=0.3,clip]{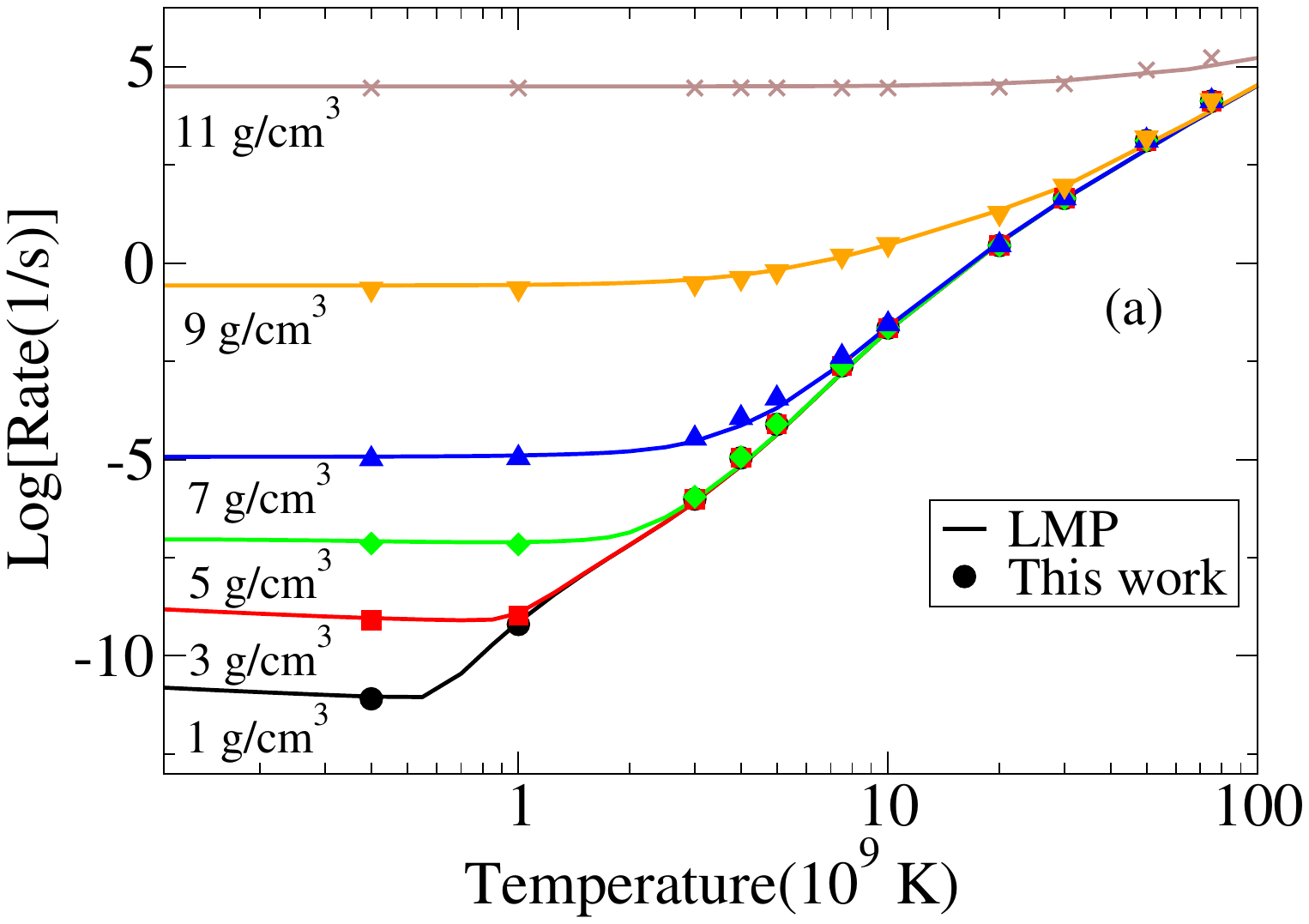}
       \includegraphics[scale=0.3,clip]{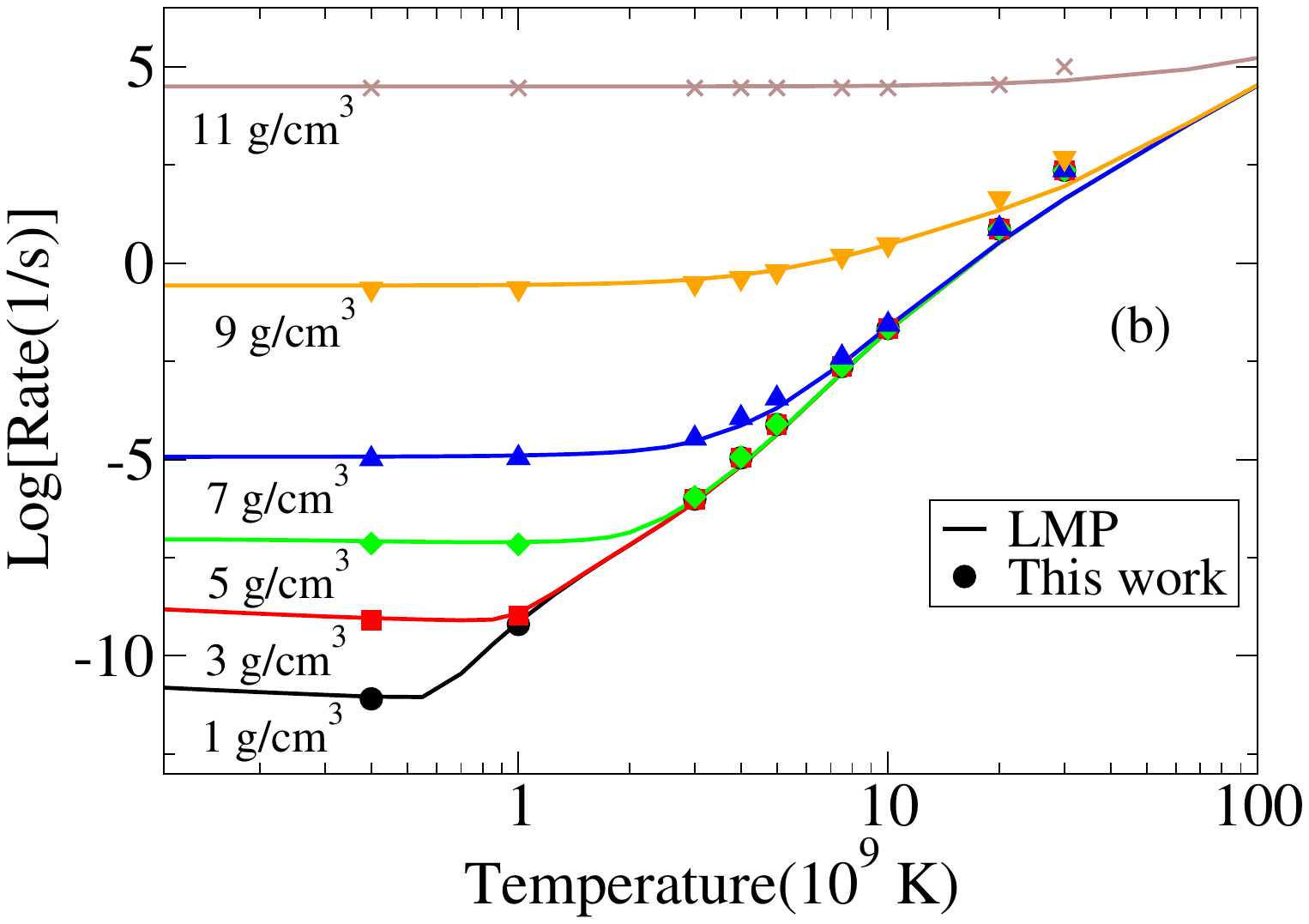}
\end{tabular}
    \caption{Thermal electron capture rates for $^{57}\mathrm{Co} \rightarrow\, ^{57}\mathrm{Fe}$ using (a) the Gaussian shell model level density, and (b)  the backshifted Fermi gas level density. Solid lines are for 
    LMP~\cite{LMP,sullivan2015sensitivity} while symbols are for this work.
    }
    \label{fig:EC_RHvLMP}
\end{figure}

 We compute the  rates for a range of temperatures and, in the case of electron capture, densities, using 
the two different models for the level density introduced in Section \ref{background}: a Gaussian distribution which closely follows the shell model level density, and which 
therefore implicitly excludes intruder states; and the commonly used backshifted Fermi gas level density. 
Yet the latter is inconsistent with our model space: 
because the latter rises exponentially with $\sqrt{E}$,  at temperatures beyond $\approx 30 \times 10^9$ K, we would 
need strength functions at initial excitation energies at well over 130 MeV. At that point we simply run out 
of states in our model space, and so we do not carry out calculations with the Fermi gas level density at the highest temperatures, i.e., $> 30 \times 10^9$K.

From 4 MeV up to 80 MeV, we used semi-converged states at 2 MeV intervals.  Although we saw evidence for a slight but systematic 
differences between transitions strength functions from initial states with different angular momenta, 
we approximated them as being identical. 
We did not investigate in detail the dependence upon the initial isospin $T_i$ (not to be confused with temperature); we expect the impact of such differences to 
be small. For example, in this case study the low-lying initial states have $T_i=3/2.$
The contributions of initial $T_i=5/2$ states are less than ten percent up to 25 MeV in initial excitation energy. 
Around 80 MeV in initial energy, inclusion of both $T_i=5/2$ and $T_i=7/2$ initial states contribute 30 percent. We note that 
any resolution in the uncertainty due to isospin dependence of strength functions at such high energies would also 
have to resolve the issue of the choice of level densities.



Fig.~\ref{fig:EC_RHvLMP} gives our electron capture rates in comparison with those of Langanke and Martinez-Pinedo \cite{LMP,sullivan2015sensitivity}, using the two different models for level densities for our calculations. 
While phase space dominates the overall behavior, 
leading to rates changing many 
orders of magnitudes, there are nonetheless differences between 
calculations. 
We see 
general agreement, although our rates are higher at high temperatures.  
We do not show the FFN~\cite{FFN3} electron capture rates, which 
are generally higher than the LMP rates, in some case by many orders of magnitude, but become much closer at high temperature and/or at high density, a general trend for nuclides in this region~\cite{LMP}.


At low temperatures and densities our electron capture results are identical to the LMP 
rates, as they must, dominated by experimental data. At higher temperatures and 
densities we found small differences from LMP, but keep in mind we used a 
different shell model interaction (GX1A). At the highest temperatures 
our results were greater than the LMP results and show an accelerating trend, but are within a factor of 2.

\begin{figure}
\begin{tabular}{cc}
    \includegraphics[scale=0.31,clip]{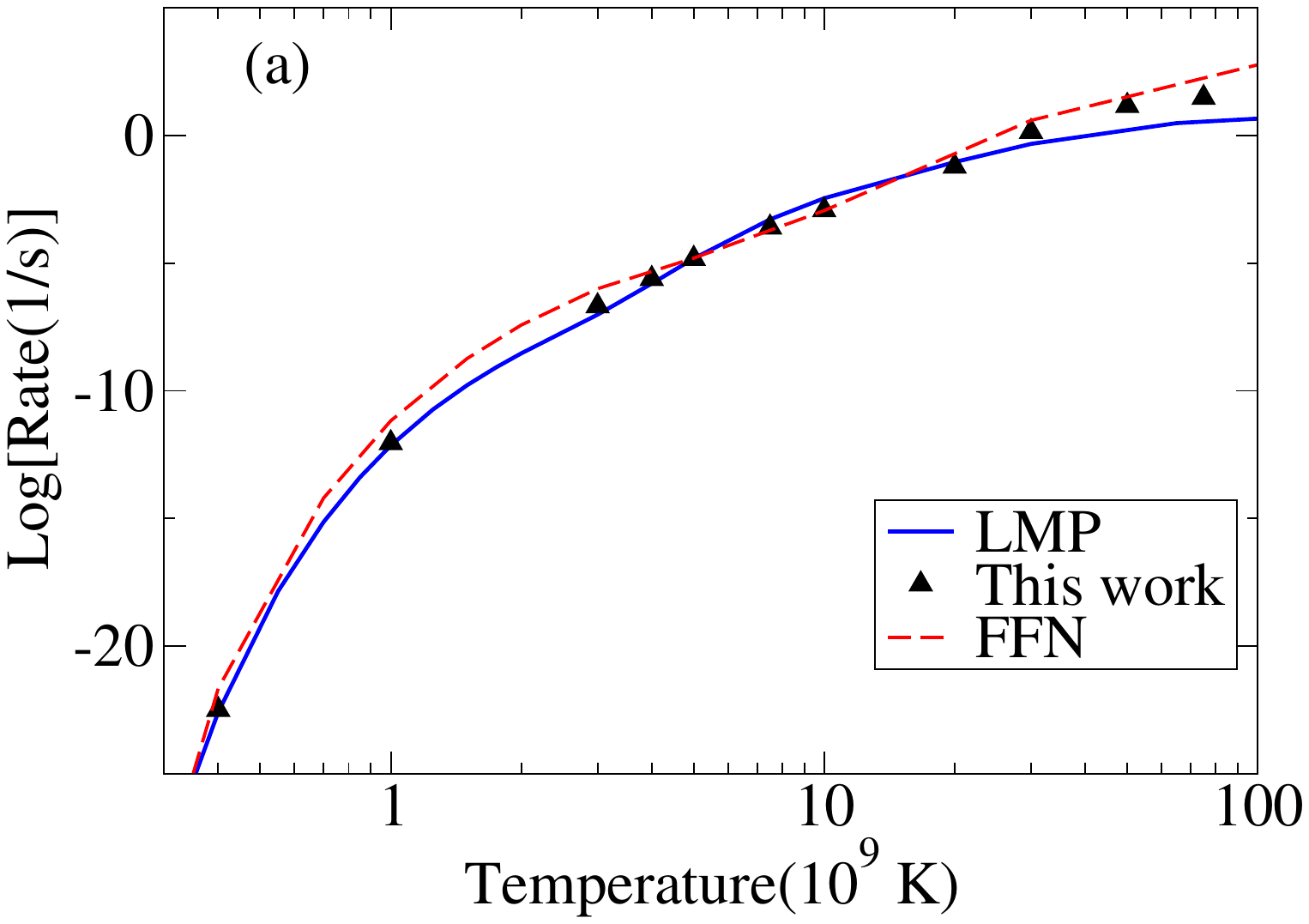}
        \includegraphics[scale=0.31,clip]{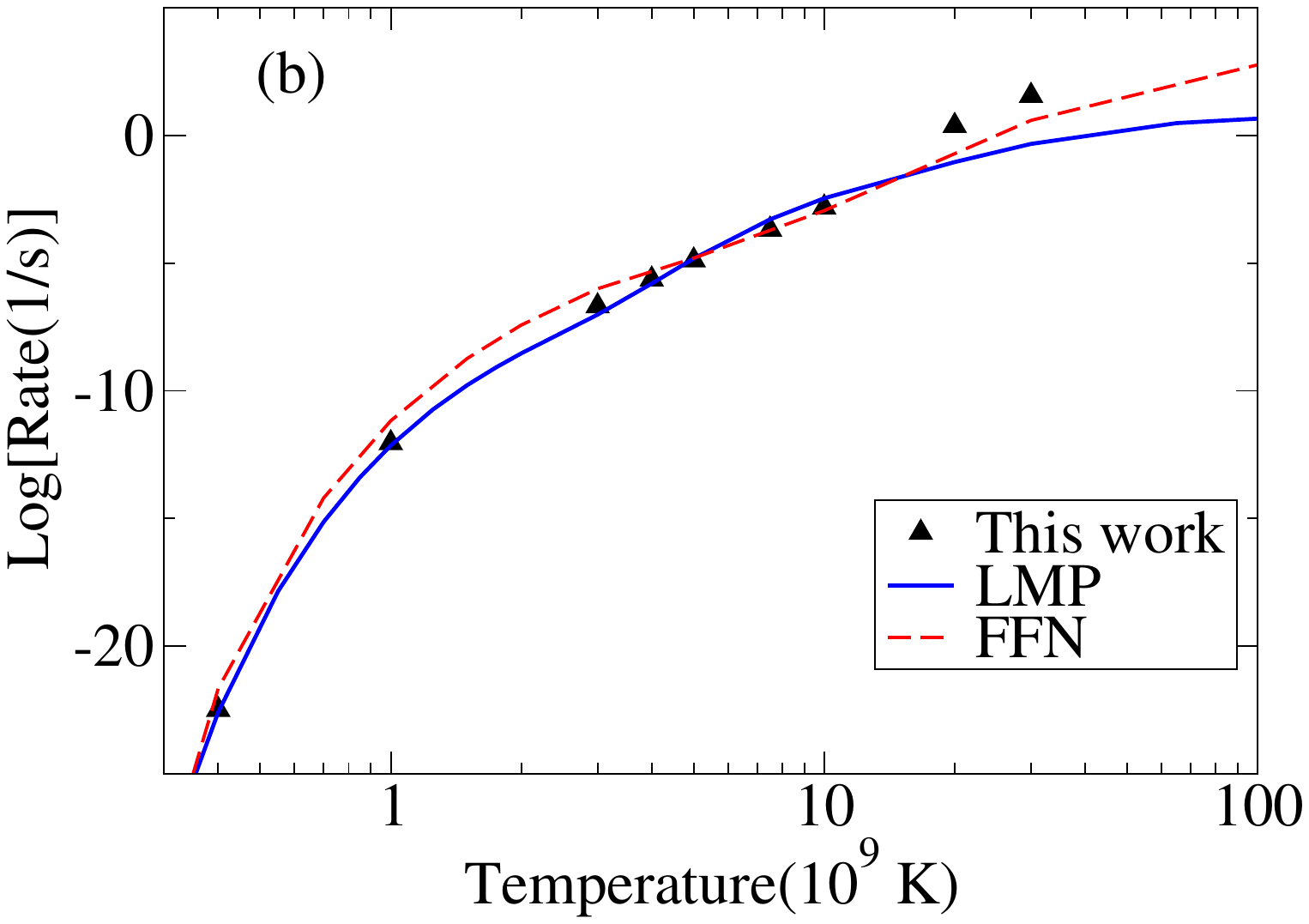}
        \end{tabular}
    \caption{Thermal positron emission rates for $^{57}\mathrm{Co} \rightarrow\, ^{57}\mathrm{Fe}$ using (a) the gaussian shell model level density, and (b)  the backshifted Fermi gas level density. The solid lines are for LMP~\cite{LMP}, dashed lines are FFN~\cite{FFN3}, and the triangles 
    are the current work. 
    }
    \label{fig:PDall}
\end{figure}

The $\beta^+$ (positron emission) rates, Fig.~\ref{fig:PDall}, show more drastic differences. The chemical potential for positrons is negligible, as is the dependence on density $\mathrm{Log}[\rho y_e ]$.  At low temperatures, the positron emission rates are many orders of magnitude smaller than electron capture; at high temperatures. however, they become comparable to other processes and may contribute in a significant way. 
As with the electron captures rates, the positron emission rates computed in this work are noticeably higher for high temperature when compared to the LMP rates and closer to or even exceed the FFN rates. Notably, the results at high temperature depend upon our choice of level density model with differences of up to a couple orders of magnitude in absolute strength.

\section{Discussion and conclusions}

In this paper, we have provided evidence that Gamow-Teller strength distributions follow an energy-localized Brink-Axel hypothesis, a hypothesis more restrictive than the generalized Brink-Axel hypothesis, but nonetheless still quite useful: strength functions from initial states nearby in energy 
are similar within statistical fluctuations. 
Because of this similarity, we can use initial states that are  semi-converged (linear combinations of eigenstates nearby in energy) 
as reasonable proxies for fully converged initial states. This allows us to obtain strength functions at higher energies than would 
otherwise be possible, up to 80 MeV in excitation energy in this work. Thus, this implies that one can still obtain strength distribution information at higher energies that would be prohibitive otherwise if one required convergence of eigenstates. 
In an application to a example system of $^{57}$Co$\rightarrow ^{57}$Fe, we generally found 
at high temperatures systematically higher 
rates than Langanke and Martinez Pinedo~\cite{LMP}, probably due to the systematic increase in total Gamow-Teller 
transition strength as one goes up in energy. {Because strength function moments behave similarly across nuclides~\cite{SumRulesBrink}, it is reasonable to anticipate similar results for 
other cases. Nonetheless, further studies to confirm should be performed, in particular to investigate if Brink hypothesis violation has systematically different effects on $\beta$-decay versus electron capture.}

This approach does not solve all problems. In any calculation in a final model space, the exhaustion of states 
must necessarily impact the centroid (energy-weighted averaged) of the transition strength function.  {While the ideal solution would be to increase the model space, 
this is not always practical in shell model calculations.  (This becomes particularly problematic as the pre-supernova core evolves to heavier and more neutron-rich isotopes~\cite{FFNII}, in which 
case the $0g_{9/2}$ orbit should be included.)} 
Thus at high energies one must either use these limited strength functions or fall back upon the standard Brink-Axel 
hypothesis, which we (and others) have shown is flawed, in conjunction with some assumption 
about the level density at high energy; {alternately, one can turn to QRPA-based 
calculations~\cite{nabi2004microscopic,nabi2005gamow,PhysRevC.76.055803,PhysRevC.81.015804,PhysRevC.100.025801,PhysRevC.101.025805}. Finally, we have not yet investigated  forbidden transitions; while these are generally much slower, in cases where allowed transitions are `blocked' forbidden transitions can compete~\cite{Fuller1982a,CoopWambach1984,PhysRevC.101.025805}.}

Nonetheless, with this approach we can tackle  transition rates at temperatures above $10^{10}$ K, which could have significant implications for neutrino production, core temperature and entropy in pre-supernova massive stars~\cite{LMP}. 
Work is ongoing to apply these methods to other $pf$-shell nuclides.

\section{Acknowledgements}

We gratefully acknowledge Wick Haxton and G. Wendell Misch for discussions and advice on various aspects of this project. This material is based upon work supported by the U.S. Department of Energy, Office of Science, Office of Nuclear Physics, under Award Number  DE-FG02-03ER41272. Also this work used the Extreme Science and Engineering Discovery Environment (XSEDE), which is supported by National Science Foundation grant number ACI-1548562, where this project was given an allocation of supercomputing resources under project ID: PHY170054. Through XSEDE, we primarily used the COMET supercomputer at the San Diego Supercomputing Center (SDSC) at University of California, San Diego (UCSD). G.M.F. acknowledges NSF Grant No. PHY-1914242 at UCSD and the NSF N3AS Physics Frontier Center, NSF Grant No. PHY-2020275, and the Heising-Simons Foundation (2017-228).
Lastly, this work was also supported by the U.S. Department of Energy, Office of Science, Office of Workforce Development for Teachers and Scientists, Office of Science Graduate Student Research (SCGSR) program. The SCGSR program is administered by the Oak Ridge Institute for Science and Education for the DOE under contract number DE‐SC0014664.

\bigskip

\appendix

\section{The Lanczos algorithm}

Here we describe briefly applications of the Lanczos algorithm \cite{cullum2002lanczos,Whitehead,ca05}, 
highlighting the key ideas in order to 
present the context for our innovations. 
The first application is well known: starting from an initial vector or
 ``pivot,'' $| v_1 \rangle $, one repeatedly applies the Hamiltonian matrix
to create a Krylov subspace:
\begin{eqnarray}
\mathbf{H} | v_1 \rangle   = & \alpha_1 | v_1 \rangle  + &\beta_1 | v_2 \rangle  \\
\mathbf{H} | v_2 \rangle  = & \beta_1 | v_1 \rangle  + & \alpha_2 | v_2 \rangle  + \beta_2 | v_3  \rangle \nonumber \\
\mathbf{H} | v_3 \rangle  = &                        & \beta_2 | v_2 \rangle  + \alpha_3 | v_3 \rangle  + \beta_3 | v_4 \rangle \nonumber \\
\ldots & & \nonumber
\end{eqnarray}
The tridiagonal matrix formed by $\alpha_i, \beta_i$, is just 
the representation of the Hamiltonian in a new orthonormal basis 
defined by the Lanczos vectors $\{ | v_i \rangle \}$.  Let $\mathbf{T}^{(k)}$ be this representation, 
truncated at $k$ vectors, that is, a $k \times k$ submatrix, achieved with $k$ iterations.  The extremal eigenpairs of $\mathbf{T}_k$ 
converge to extremal eigenpairs of $\mathbf{H}$ as $k$ increases.  Although it depends upon 
the system, and specifically on the density of states, one typically obtains  5 to 10 converged eigenpairs 
with $k$ somewhere between 100 and 300; 150 converged eigenpairs can require thousands of iterations, usually 
requiring the thick-restart Lanczos method~\cite{wu2000thick} which saves on storage of vectors and on reorthogonalization.  One can understand the 
effectiveness of the Lanczos algorithm in terms of moments \cite{whitehead1978lanczos,whitehead1980moment}.

\begin{figure}
    \centering
    \includegraphics[width=0.7\textwidth]{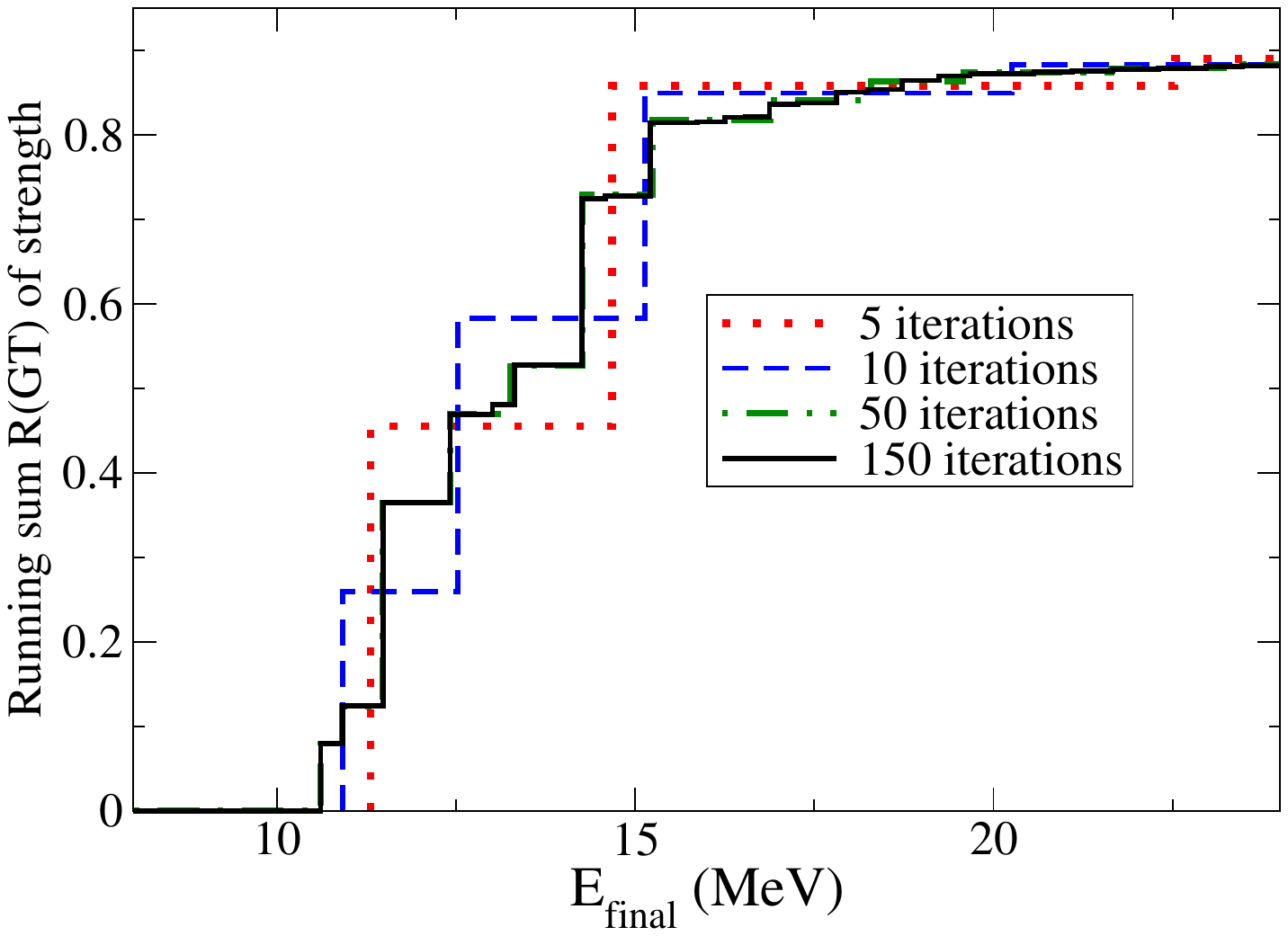}
    \caption{Convergence of the running sum of the strength function for the transition $^{28}\mathrm{Si} \rightarrow \, ^{28}\mathrm{Al}$ as a function of final state energy, computed in the $sd$-shell with the USDB interaction~\cite{PhysRevC.74.034315}.}
    \label{fig:strfunc}
\end{figure}

The second application is also well-known to practitioners \cite{ca05,whitehead1980shell,bloom1984gamow}. Suppose one 
wants to obtain the strength function of some transition operator $\hat{\cal O}$, in our case the Gamow-Teller operator, from some initial 
state $| \psi_0\rangle $ previously calculated. To do this, choose the pivot to be 
$ |v_1 \rangle = {\cal N}^{-1/2} \hat{\cal O} | \psi_0 \rangle $, where the normalization
${\cal N} =  \langle \psi_0 |\hat{\cal O}^\dagger \hat{ \cal O} | \psi_0 \rangle$ is  the total strength. One 
then carries out the Lanczos algorithm in the usual way.  Let the eigenvectors of $\mathbf{T}^{(k)}$ be 
given defined by the columns of the matrix $\mathbf{L}$, that is, let
\begin{equation}
\sum_j {T}^{(k)}_{nj}L_{jf} = \tilde{E}_f L_{nf}     
\end{equation}
where $\tilde{E}_f$ is the $f$th eigenvalue of $\mathbf{T}^{(k)}$ and may be unconverged. One can transform 
these eigenvectors back to the original space:
\begin{equation}
| \tilde{\psi}_f \rangle = \sum_n | v_n \rangle L_{nf},    
\end{equation}
where we use the notation $\tilde{\psi}$ to denote states that may not be fully converged. 
It is easy to see that $\langle v_1 | \tilde{\psi}_f \rangle = L_{1f}$, but 
the transition strength from $\psi_0$ to $\tilde{\psi}_f$ is  
$|\langle \tilde{\psi}_f | \hat{\cal O} | \psi_0 \rangle |^2 ={\cal N} | \langle \tilde{\psi}_f | v_1 \rangle |^2 $,
but $\langle v_1 | \tilde{\psi}_f \rangle = L_{1f}$, so that 
$|\langle \tilde{\psi}_f | \hat{\cal O} | \psi_0 \rangle |^2= {\cal N} | L_{1f}|^2$. In other words, one can easily read off the strength from the eigenvectors in the space of 
Lanczos vectors. Even for unconverged states, this information is useful, as it reproduces the moments
of the strength function, which in turn means even a few tens of Lanczos iterations produces extremely accurate 
representations of the strength function.  We illustrate this in Fig.~\ref{fig:strfunc}, where we computed the running sum 
of Gamow-Teller strengths from the ground state of $^{28}$Si $\rightarrow ^{28}$Al; the Hamiltonian is the universal 
$sd$-shell effective interaction version `B,' or USDB~\cite{PhysRev.74.1046}. Even with only five Lanczos iterations the 
base outline of the strength function is established, and the difference between 50 iterations and 150 iterations is 
nearly indistinguishable. 

In order to get all the correct factors, one needs to project out 
states of good angular momentum and, often, isospin. This projection or decomposition can 
also be accomplished by the Lanczos algorithm \cite{morrison1974novel,PhysRevC.91.034313,PhysRevC.95.024303}.  
Suppose one has a state $ | \psi \rangle = \sum_a c_a | \omega_a \rangle$,
where $| \omega_a \rangle$ are eigenstates of a scalar operator, typically a Casimir operator such as $\hat{J}^2$.
To extract the $ | \omega_a \rangle$, just use $| \psi \rangle$ as the pivot. Carrying out Lanczos with 
the Casimir operator produces the projection of the pivot $| \psi \rangle$ in the subspaces defined by the Casimir operator. In most cases of interest the expansion has only a very few terms, requiring correspondingly very few Lanczos iterations. 

Finally, we developed a novel method to obtain semi-converged excited states, 
also called ``interior'' eigenpairs.  The problem of finding interior 
eigenpairs of very large matrices is a notoriously difficult problem. 
After experimenting with different approaches, we use a modified version 
of the ``thick-restart'' Lanczos algorithm \cite{wu2000thick}, which is also widely used.

In the thick-restart Lanczos algorithm, after generating a fixed number $k$ 
of Lanczos vectors, one diagonalizes $\mathbf{T}^{(k)}$. One then restarts 
the process, keeping $N_\mathrm{thick}$ of the lowest eigenpairs and, through 
repeated application of $\mathbf{H}$, obtains $k-N_\mathrm{thick}$ additional 
vectors. This method is useful when the dimensionality is very large and 
one cannot store many Lanczos vectors, or if one wants to avoid the 
numerically necessary reorthogonalization that eventually dominates 
the Lanczos algorithm if one otherwise goes to very large $k$.  The 
thick-restart algorithm has also been extended to block Lanczos \cite{shimizu2019thick}.

Rather than keeping the lowest eigenpairs, we choose a target energy 
$E_\mathrm{target}$ and keep nearby eigenpairs, that is, 
we take the $N_\mathrm{thick}$ lowest pairs defined by $| E_i - E_\mathrm{target}|$. 
This approach converges slowly, in part due to the very high density of 
eigenvalues.  (Note: we tried alternatives, such as diagonalizing 
$(\mathbf{H}-E_0)^2$, but such alternatives performed no better and 
often worse than our chosen methodology.) Because of the localized Brink-Axel hypothesis, however, we do not 
need fully converged states; all we need are states relatively localized in energy.

\bibliographystyle{apsrev4-1}
\bibliography{RHmaster}
 \end{document}